\pdfoutput=1
\documentclass[sigconf,nonacm]{acmart}

\usepackage{enumitem}   
\usepackage{float}
\usepackage{multirow}
\usepackage{tabularx}
\usepackage{tikz}
\usetikzlibrary{arrows.meta, positioning, shapes.geometric, fit, calc}
\usepackage{algorithm}
\usepackage{algpseudocode}
\usepackage{subfiles}   
\usepackage{graphicx}
\usepackage[normalem]{ulem} 

\setcopyright{none}
\acmYear{2027}
\acmConference[KDD '27]{ACM KDD 2027}{August 1--5, 2027}{San Jose, CA, USA}
\acmDOI{}
\acmISBN{}

\newcommand{\todo}[1]{}
\newcommand{\note}[1]{}
\newcommand{\miley}[1]{}
\newcommand{\sophia}[1]{}

\newif\ifshowgeshi
\showgeshifalse  
\newcommand{\geshi}[1]{\ifshowgeshi\textcolor{teal}{\textbf{[GS: #1]}}\fi}
\newcommand{\gsdone}[1]{\ifshowgeshi\textcolor{green!45!black}{\textbf{[GS done: #1]}}\fi}

\providecommand{\shortauthors}{}

\begin{document}
\title{Measuring Public Participation in Local Government: An Audio-Based Instrument for Classifying Speakers in City Council Meetings}

\author{Menglin (Miley) Liu}
\affiliation{%
  \department{Division of Computational Social Science, School of Humanities and Social Science}
  \institution{The Chinese University of Hong Kong, Shenzhen}
  \city{Shenzhen}
  \state{Guangdong}
  \country{China}}
\email{menglinliu@cuhk.edu.cn}

\author{J. Sophia Wang}
\orcid{0000-0002-8474-4001}
\affiliation{%
  \department{Department of Political Science}
  \institution{Yale University}
  \city{New Haven}
  \state{Connecticut}
  \country{USA}}
\email{jinghong.wang@yale.edu}

\author{Ge Shi}
\orcid{0000-0002-6944-6874}
\affiliation{%
  \department{Department of Computer Science}
  \institution{University of California, Davis}
  \city{Davis}
  \state{California}
  \country{USA}}
\email{geshi@ucdavis.edu}

\renewcommand{\shortauthors}{Liu, Wang, and Shi}

\begin{abstract}
How much does the public actually speak in local government? Despite sustained interest in democratic participation, we lack systematic, large-scale measures of public voice at the local level, because the underlying record---audio of open meetings---has been prohibitively costly to convert into data. We contribute a scientific measurement instrument that addresses this barrier: an audio-analysis pipeline that classifies each speaker in a city council meeting as an elected official or a member of the public directly from the audio signal, and aggregates these labels into validated measures of participation such as floor-time share. The instrument first segments each recording by speaker---identifying who speaks when---and then classifies each speaker with a multimodal large language model (MLLM), using the meeting's own temporal structure (neighboring speakers, iterative refinement, and a majority vote across three classifiers) as supporting components. We collect 1,618 recordings from seven U.S.\ cities (2017--2023); after diarization and quality control, the analytic corpus is 1,565 meetings from five cities. The task is non-trivial: audio quality is uneven, members of the public speak informally and are often introduced only briefly, and speech frequently overlaps. Even so, human validation shows the segmentation stage achieves 89\% accuracy with high inter-rater reliability (mean pairwise $\kappa = 0.84$; range $0.69$--$1.00$), and the classifier reaches 87.7\% accuracy ($\kappa = 0.736$, public-participant F1 $= 0.832$) on ground truth from five cities ($N = 1{,}761$ speakers across eight annotation rounds)---agreement with the human consensus comparable to the agreement among the annotators themselves. Applying the instrument at scale, we document that members of the public receive a median of just 10\% of meeting floor time and that the top speaker is an official in over 94\% of meetings; we also describe a decline in public participation over the COVID-19 period that had not recovered by 2023. The instrument measures democratic participation directly from meeting audio, without transcription or linkage to external voter or property records, establishing audio-as-data as a practical approach for the study of local politics.
\gsdone{Title/abstract consistency addressed: the abstract now distinguishes the 1,618 collected recordings from the 1,565-meeting analytic corpus used for the participation results.}
\end{abstract}



\maketitle

\geshi{Submission note: the main paper exceeds eight pages. Keep every required section within pages 1--8, followed by references and a reproducibility-only appendix (Action~8, still open).}

\gsdone{Action 1---Contribution framing: position the primary contribution as an AI-enabled scientific measurement instrument that converts public-meeting audio into validated measures of participation. Present diarization, temporal exemplars, iterative refinement, and voting as supporting components rather than a new general-purpose ``multi-agent'' architecture. Organize the claims as measurement, methodology, validation, domain-science finding, and---only if artifacts are released---resource contributions.}

\gsdone{Action 2---Scientific claims and terminology: center one defensible political-science finding enabled by the instrument. Treat the cross-city and post-2020 patterns as descriptive unless a causal design is added; soften ``first,'' ``confirms,'' ``practical ceiling,'' ``persistent structural shift,'' and similar claims. Reconsider whether the operational label should be ``public participant'' rather than ``citizen,'' because legal citizenship is not observed.}

\gsdone{Action 3---Evaluation validity: add one explicit table separating prompt development, exemplar selection, ensemble/pass selection, final evaluation, and production data, including whether speakers or meetings overlap. Avoid saying there is no train/test split when prompts and operating choices were empirically selected. Validate a subset against independent sources such as rosters, agendas, minutes, or speaker cards, and justify Pass~3 despite Pass~2 having higher accuracy and F1.}

\gsdone{Action 4---Measurement uncertainty: supplement pooled speaker-level correction with meeting-clustered uncertainty, city/period sensitivity, duration-weighted error analysis for floor share, and uncertainty intervals on the principal application figures. State the assumptions required to transport sensitivity and specificity from the labeled sample to cities and periods with different class prevalence and audio conditions.}

\gsdone{Action 5---Reproducibility, ethics, and AI disclosure: report the exact Gemini identifier, API dates, settings, exemplars, seeds, retries, failures, safety filters, unparsable-response handling, cost, repeated-run stability, and artifact location. Clarify IRB or human-subjects determination, consent expectations for public recordings, third-party retention terms, voice-data storage/access, demographic and accent-related risks, and redistribution limits. In Generative AI Usage, name every tool and its role in classification, code, analysis, literature work, drafting, and proofreading.}

\gsdone{Action 6---Writing and section focus: keep most paragraphs to one claim in roughly 80--150 words using claim--evidence--interpretation/limitation. Structure the Introduction around the scientific question, measurement barrier, AI opportunity, approach/validation, and contributions; expand Related Work into political participation, computational measurement, audio/diarization/MLLM work, and the unresolved gap. In Method, emphasize scientific necessity over API mechanics; in Results, give each paragraph one headline result, comparison/uncertainty, interpretation, and limitation. Remove repeated metrics, ``without transcription'' claims, promotional wording, and reader-directed phrases such as ``the reader does not need.''}

\gsdone{Action 7---Corpus documentation: add a compact Corpus Documentation or Data Statement table covering purpose, sources, dates, collection frame versus analytic sample, exclusions, scale, units, formats, preprocessing, annotations, data partitions, leakage controls, missingness, biases, access, licensing, privacy, intended/prohibited uses, versioning, maintenance, and citation. A formal data card or Croissant file is not required for AI for Sciences; Croissant is useful only if a stable ML-ready derived corpus is released. A repository should provide source manifests, retrieval/preprocessing code, releasable annotations and metadata, split files, prompts/exemplars, responsible-use terms, and a versioned DOI where possible.}

\geshi{Action 8---Eight-page scientific narrative: ensure the mandatory Limitations and Ethical Considerations and Generative AI Usage sections appear within pages 1--8, followed immediately by references. Keep essential evidence---cross-city generalization, statistical comparisons, measurement-error implications, and the audio-versus-text result---in compact form in the main paper; reserve the appendix for reproducibility details, proofs, prompts, and pseudocode. Remove all GS editing notes only when preparing the final submission PDF.}

\gsdone{Action 9---Bibliography verification: replace manually constructed or incomplete entries in \texttt{references.bib} with standard exported citations. For each cited work, search for the authoritative publisher page or official arXiv record, use its ``Cite'' function, select BibTeX, and copy the exported entry into \texttt{references.bib}. Then preserve the paper's existing citation key where necessary, check authors, title, venue, year, volume, issue, pages, DOI, and URL, remove duplicate or nonstandard fields, and rebuild to confirm that every citation resolves correctly.}


\section{Introduction}
\label{sec:intro}

Local government is where most Americans encounter the state directly. City council meetings---open forums where residents can address their elected representatives---are a fundamental channel of democratic participation. Yet we know remarkably little about who actually speaks in these meetings, for how long, and how participation patterns vary across cities, time periods, and institutional contexts. Theories of participatory and deliberative democracy hold that legitimate collective decisions require the direct involvement of those they affect, not merely periodic elections \citep{pateman1970participation, mansbridge1983beyond, fung2004empowered}, and public comment is one of the few institutionalized channels through which residents voice preferences to officials between elections.

The value of these forums is contested. A familiar critique, in both scholarship and popular culture, is that the open microphone is captured by a vocal and unrepresentative minority: those with the time, confidence, and resources to attend dominate the floor, amplifying narrow grievances while the broader public goes unheard \citep{einstein2019still, einstein2020neighborhood}. Whether public comment broadens democratic voice or narrows it is, at bottom, an empirical question about who holds the floor and for how long. Answering it at scale would let researchers ask how institutional design shapes the distribution of voice between the public and their representatives---and, once speaking-time measures are linked to external data, whether some groups are heard more than others. These are first-order questions for the study of local democracy, yet they remain out of reach. Our instrument targets the prior, measurable question---how much of the floor the public holds relative to officials---and, as we detail below, observes neither the demographics nor the identities of speakers; the compositional questions are enabled by, but not answered in, this paper.

The obstacle is a measurement barrier. Existing work on political participation has focused overwhelmingly on voting, campaign contributions, and contacting officials \citep[e.g.,][]{verba1995voice, brady1995beyond}. A growing body of work examines public comment in local government \citep{einstein2019still, sahn2024public}, but it relies on labor-intensive manual coding of meeting minutes or transcripts, which limits it to small numbers of meetings. A single meeting can last two to three hours and feature dozens of speakers; transcribing and coding even one requires hours of skilled labor. As a result, systematic data on public participation in local government remains sparse, anecdotal, or confined to narrow case studies \citep{einstein2019still, einstein2020neighborhood}. The record itself is abundant---thousands of meetings are posted publicly as audio and video---but it has been prohibitively costly to convert into measures.

Audio-native multimodal models change what is measurable. This paper develops a measurement instrument that converts public-meeting audio into validated measures of participation. The instrument combines neural speaker diarization (identifying \emph{who} speaks \emph{when}) with a multimodal large language model (MLLM) that classifies each speaker as an official or a member of the public directly from the audio signal, exploiting the meeting's own temporal structure to resolve ambiguous cases. We validate every stage against human annotation and propagate the residual classification error into the reported measures. Applied to 1,618 meetings from seven U.S.\ cities between 2017 and 2023, the instrument yields the first speaking-time measures of public participation at this scale.

Our contributions are organized around the instrument. \textbf{A new measurement target and problem setting:} we measure public participation as \emph{floor-time share}---not merely headcount---from meeting audio alone, without transcription or linkage to external voter or property records, and we do so in a setting prior audio-as-data work has not addressed: no speaker roster, no identity labels, and no known number of speakers $K$ are given in advance (prior work classifies speakers whose identities are known a priori). \textbf{A label-free method that exploits meeting structure:} the technical insight is that a meeting's own temporal structure can be turned into a classification signal---through temporal exemplar selection, iterative self-refinement that bootstraps from the model's \emph{own} predictions rather than any labeled training data, and a majority vote across complementary classifiers---raising accuracy over a strong single-pass audio baseline with no task-specific training. \textbf{Generalization without retraining:} because the pipeline trains no task-specific weights, it transfers to new jurisdictions off the shelf; leave-one-city-out evaluation shows it holds accuracy on held-out cities where supervised acoustic baselines degrade sharply---the property that makes measurement at national scale practical, and one that per-setting fine-tuning would forfeit. \textbf{Validation:} we validate diarization and classification against human ground truth from five cities, report agreement that approaches the agreement among human annotators themselves, evaluate on a held-out test set, and carry the resulting measurement error through to the substantive estimates. \textbf{Domain finding:} using the instrument, we document that members of the public receive far less floor time than their headcount would suggest, that speaking time is highly concentrated among officials, and that public participation declined over the COVID-19 period. We release the pipeline as open-source software so that the same measurement strategy can be applied to other deliberative settings.

Section~\ref{sec:background} reviews the literature on local participation and its measurement. Section~\ref{sec:data} describes the corpus. Section~\ref{sec:method} presents the instrument. Section~\ref{sec:validation} reports validation. Section~\ref{sec:results} presents the substantive findings. Section~\ref{sec:discussion} discusses implications and limitations.


\section{Related Work}
\label{sec:background}

\noindent\textbf{Political participation and public comment.} Research on political participation has centered on voting, contributions, and contacting officials, and on the resource and mobilization gradients that structure them \citep{verba1995voice, brady1995beyond}. A more recent strand turns to local public comment as a distinct arena, showing that those who show up are older, whiter, and more likely to be homeowners than the populations they claim to speak for \citep{einstein2019still, einstein2020neighborhood, sahn2024public}. This work establishes \emph{who participates} but measures it through sign-in sheets and hand-coded minutes, which caps sample sizes at tens of meetings and records presence rather than the amount of floor time each speaker receives.

\noindent\textbf{Computational measurement of political speech.} The dominant computational paradigm, Text-as-Data \citep{grimmer2013text}, treats transcripts as the unit of analysis. Applied to spoken proceedings it requires an upstream transcription step, which is costly at scale and inherits the documented demographic bias of automatic speech recognition (ASR) against non-standard dialects \citep{koenecke2020racial}. Recent local-politics pipelines adopt exactly this architecture---ASR and diarization as infrastructure that produces text for downstream coding \citep{martin2026participation, xu2025publicspeak, barari2023localview}---and in doing so discard the acoustic and turn-taking signal that the audio itself carries.

\noindent\textbf{Audio-as-data, diarization, and multimodal models.} A parallel ``Audio-as-Data'' line analyzes the speech signal directly---pitch, emotional arousal, and prosody \citep{dietrich2019pitch, knox2021dynamic, mestre2025potential}---but studies elite speakers (legislators, judges, candidates) whose identities are known in advance, so it never confronts the problem of deciding \emph{who} an unlabeled speaker is. The enabling technologies for that problem have matured: neural speaker diarization now segments ``who spoke when'' from raw audio with a single open-source model \citep{bredin2023pyannote, park2022review}, and audio-native multimodal LLMs can reason over the speech signal without an intermediate transcript \citep{gemini2024}. Neither has been applied to measuring mass participation.

\noindent\textbf{The gap.} No existing approach measures how much ordinary members of the public actually speak in local government at scale. Participation scholarship has the substantive question but not the measurement technology; audio-as-data has the signal but studies elites; text-as-data pipelines transcribe away the very cues that distinguish an official from a member of the public. We close this gap with an instrument that classifies speaker roles directly from audio using a multimodal LLM---no transcription, no linkage to external voter or property records---and exploits meeting structure through iterative refinement and a sliding-window ensemble to turn role labels into validated participation measures.

\section{Data}
\label{sec:data}

We study seven U.S.\ cities between 2017 and 2023. The sample is purposive rather than representative: cities were selected for publicly available YouTube recordings of council meetings, geographic spread across three states (California, Massachusetts, and New Jersey), variation in council size and meeting format, and sufficient meeting volume for longitudinal analysis. The cities were originally drawn from a sampling frame of municipalities with rent control policies, though the substantive analysis of housing policy is not the focus of this paper. Table~\ref{tab:cities} summarizes the downloaded recordings.

\begin{table*}[t]
\centering
\caption{Sample Cities}
\label{tab:cities}
\begin{tabular}{llrr}
\toprule
City & State & Meetings & Years \\
\midrule
Brockton       & MA & 503 & 2017--2023 \\
Asbury Park    & NJ & 372 & 2017--2023 \\
Jersey City    & NJ & 301 & 2017--2023 \\
New Brunswick  & NJ & 277 & 2017--2023 \\
Inglewood      & CA & 154 & 2017--2023 \\
Oakland        & CA &   9 & 2017--2018 \\
East Palo Alto & CA &   2 & 2018 \\
\midrule
\textbf{Total} &    & \textbf{1,618} & \\
\bottomrule
\end{tabular}
\begin{flushleft}
\footnotesize\textit{Note:} Counts are of recordings downloaded from each city, i.e.\ the full collection frame. Oakland and East Palo Alto are listed for transparency but do not enter the analytic corpus (Table~\ref{tab:corpus_desc}); see text.
\end{flushleft}
\end{table*}

Table~\ref{tab:cities} lists every city from which we downloaded recordings, including two that do not enter the analytic corpus. Oakland and East Palo Alto contribute fewer than a dozen meetings each; Oakland is excluded from all subsequent analyses because its recordings were too short or corrupted for reliable diarization, and East Palo Alto (2 meetings) is retained for ground-truth development only. The analytic corpus is therefore the five remaining cities (Table~\ref{tab:corpus_desc}). The 51-meeting reduction from 1,618 downloaded recordings to 1,567 diarized meetings reflects additional losses from corrupted downloads, audio files shorter than the minimum duration threshold, and diarization failures across all cities.

\paragraph{Sample representativeness.} These cities are not a representative cross-section of the roughly 30,000 U.S.\ general-purpose local governments, and we make no claim that our patterns generalize to all of them. All five analytic cities are dense, urban or suburban, and in Democratic-leaning states; the rent-control sampling frame and the requirement that a city post meetings to YouTube compound this skew, favoring larger jurisdictions with professionalized clerk offices and active civic constituencies, while rural and Republican-dominated jurisdictions are absent. We treat the study as a proof of concept for the measurement instrument rather than a census of American local democracy (Section~\ref{sec:discussion}). Coverage is also uneven over time---the bulk of recordings fall in 2019--2022, with a pandemic-era surge in 2020--2021 and few Brockton recordings before 2020---which we account for in the longitudinal analysis of Section~\ref{sec:results}.

Meeting recordings were obtained from official city YouTube channels and government websites via automated scraping (Section~\ref{sec:appendix_scraping}); all are public records of open meetings, and we assign de-identified speaker labels and classify only by institutional role (Section~\ref{sec:ethics}).

After speaker diarization (described in Section~\ref{sec:method}), the corpus yields 199,118 speech segments from 29,257 unique meeting-level speakers across 1,567 meetings, totaling approximately 2,600 hours of audio. The median meeting contains 16 identified speakers (mean = 18.7) and runs 76 minutes, though meetings range from brief special sessions to multi-hour hearings. At the speaker level, each speaker produces a median of 7 speech segments (mean = 6.8). City-level variation is considerable: Jersey City meetings average 30.6 speakers per meeting, reflecting that city's large council and active public comment periods, while Brockton averages 13.2. Table~\ref{tab:corpus_desc} presents descriptive statistics by city.

\begin{table*}[t]
\centering
\caption{Corpus Descriptive Statistics by City}
\label{tab:corpus_desc}
\begin{tabular}{lrrrr}
\toprule
City & Meetings & Speakers & Segments & Avg.\ Speakers/Meeting \\
\midrule
Asbury Park    & 364 & 7,270  & 46,030  & 20.0 \\
Brockton       & 499 & 6,605  & 48,899  & 13.2 \\
Inglewood      & 141 & 2,346  & 16,869  & 16.6 \\
Jersey City    & 299 & 9,143  & 66,515  & 30.6 \\
New Brunswick  & 262 & 3,880  & 20,714  & 14.8 \\
East Palo Alto &   2 &    13  &     91  &  6.5 \\
\midrule
\textbf{Total} & \textbf{1,567} & \textbf{29,257} & \textbf{199,118} & \textbf{18.7} \\
\bottomrule
\end{tabular}
\end{table*}

A full data statement for the corpus---consolidating purpose, provenance, scale, preprocessing, partitions, leakage controls, known biases, and access terms---is provided in Appendix~\ref{sec:appendix_corpus_doc} (Table~\ref{tab:datastatement}). A formal data card or Croissant record is not required for this venue; the accompanying repository instead provides source manifests, retrieval and preprocessing code, releasable annotations and split files, and the prompts and exemplars used.

\subsection{Ground Truth Annotation}
\label{sec:ground_truth_overview}

We distinguish three data partitions. (1)~\textbf{Development data} consists of meetings from two cities not in the production sample---Albany, NY (1~meeting, 6~speakers) and Chino, CA (3~meetings, 45~speakers)---used exclusively for prompt tuning and diarization validation; no development labels enter the production evaluation. (2)~\textbf{Validation ground truth} covers the five production cities (Asbury~Park, Brockton, Inglewood, Jersey~City, and New~Brunswick) across eight annotation rounds (R2, R4--R7 and R9--R11, with R8 excluded for low inter-rater reliability). Three independent research assistants labeled each speaker as ``official'' or ``public participant'' from audio clips alone; disagreements were resolved by majority vote. The resulting validation set contains $N = 1{,}761$ speakers (1,069 officials, 692 public participants; 39.3\% public-participant rate) from 165 meetings, with mean Fleiss' $\kappa = 0.723$ (range $0.525$--$0.868$) across rounds. (3)~\textbf{Production corpus}: the 29,238 speakers classified at scale carry no human labels and are used only for the application in Section~\ref{sec:results}. The classifier requires no task-specific \emph{training}, so no speakers are used to fit model weights; but this does not mean the pipeline is free of data-dependent choices. Prompt wording, the few-shot exemplar pool, the number of bootstrapping passes, and the ensemble rule were all selected empirically on development and validation data. We therefore report these as operating choices (Section~\ref{sec:validation}, Table~\ref{tab:partitions}) rather than claiming an assumption-free ``no train/test split,'' and we keep the production corpus strictly held out from all such selection. Full annotation details, per-round reliability, and per-city counts are reported in Appendix~\ref{sec:appendix_ground_truth}.


\section{Method}
\label{sec:method}

We first state the classification problem formally, then describe the three-stage pipeline that solves it.

\textbf{Problem statement.}  Let $\mathbf{x} \in \mathbb{R}^{T}$ denote a raw audio recording of duration~$T$.  Speaker diarization partitions $\mathbf{x}$ into segments $\{(s_i, e_i, k_i)\}_{i=1}^{N}$, where $s_i$ and $e_i$ are start and end times and $k_i \in \mathcal{K} = \{1, \ldots, K\}$ is a speaker label, inducing a set of $K$ unique speakers.  The classification task is to learn a mapping $f\colon \mathcal{K} \to \{0, 1\}$, where $0 =$ \emph{official} and $1 =$ \emph{public participant} (a member of the public; we adopt this operational label rather than ``citizen'' because legal citizenship is not observed).  Given $M$ classifiers $f_1, \ldots, f_M$, we combine predictions via ensemble rules: \emph{majority vote} $\hat{y}_k = \mathbf{1}\!\bigl[\sum_{m=1}^{M} f_m(k) \geq \lceil M/2 \rceil\bigr]$; \emph{AND} $\hat{y}_k = \mathbf{1}\!\bigl[f_a(k) = 1 \;\wedge\; f_b(k) = 1\bigr]$; and \emph{OR} $\hat{y}_k = \mathbf{1}\!\bigl[f_a(k) = 1 \;\vee\; f_b(k) = 1\bigr]$.  Theoretical properties of these ensemble rules---including connections to the Condorcet jury theorem---are developed in Appendix~\ref{sec:appendix_theory}.

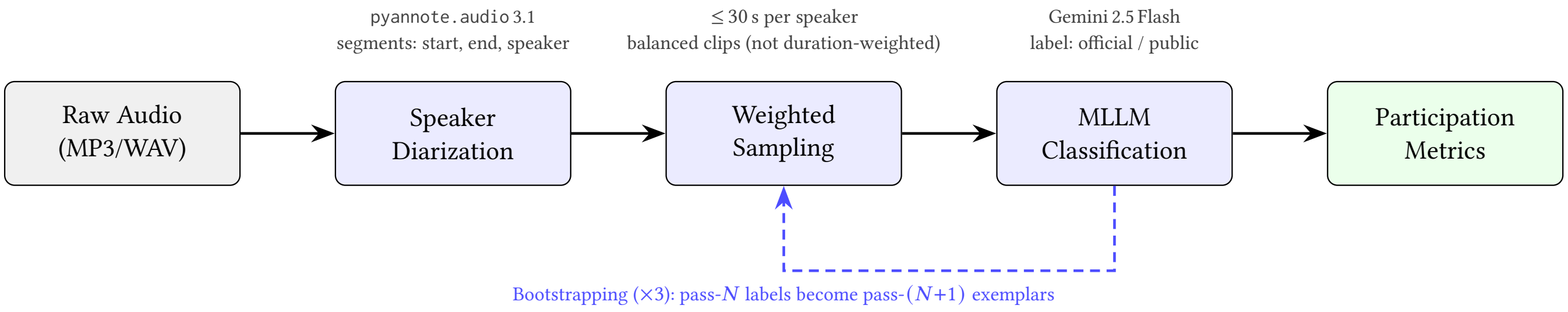
\begin{figure*}[t]
\centering
\resizebox{\linewidth}{!}{%
\begin{tikzpicture}[
    node distance=1.0cm,
    box/.style={rectangle, draw, rounded corners=3pt, minimum width=2.5cm,
                minimum height=1.1cm, align=center, font=\small},
    input/.style={box, fill=gray!12},
    stage/.style={box, fill=blue!8},
    output/.style={box, fill=green!8},
    arrow/.style={-{Stealth[length=2.5mm]}, thick},
    ann/.style={font=\scriptsize, align=center, text=black!75}
]

\node[input]  (raw)   {Raw Audio\\(MP3/WAV)};
\node[stage,  right=of raw]   (diar)  {Speaker\\Diarization};
\node[stage,  right=of diar]  (samp)  {Weighted\\Sampling};
\node[stage,  right=of samp]  (class) {MLLM\\Classification};
\node[output, right=of class] (out)   {Participation\\Metrics};

\draw[arrow] (raw)  -- (diar);
\draw[arrow] (diar) -- (samp);
\draw[arrow] (samp) -- (class);
\draw[arrow] (class) -- (out);

\node[ann, above=5pt of diar]  {\texttt{pyannote.audio}\,3.1\\[1pt] segments: start, end, speaker};
\node[ann, above=5pt of samp]  {$\le\!30$\,s per speaker\\[1pt] balanced clips (not duration-weighted)};
\node[ann, above=5pt of class] {Gemini\,2.5\,Flash\\[1pt] label: official / public};

\draw[arrow, densely dashed, blue!70]
    (class.south) -- ++(0,-0.9) -| (samp.south)
    node[ann, text=blue!70, pos=0.5, below=1pt]
        {Bootstrapping ($\times 3$): pass-$N$ labels become pass-$(N{+}1)$ exemplars};

\end{tikzpicture}
}
\caption{Audio-analysis pipeline. Raw meeting audio is diarized into per-speaker
segments (each with a start time, end time, and speaker index); weighted sampling
then draws balanced per-speaker clips---capped at 30\,s each---so brief public
speakers are not crowded out by long-speaking officials; and a multimodal LLM
classifies each speaker as an official or a member of the public. The dashed loop is
the bootstrapping refinement: each pass's predicted labels supply the few-shot
exemplars for the next, over three passes (Section~\ref{sec:bootstrapping}). Per-speaker
labels are then aggregated into participation metrics such as floor-time share.}
\label{fig:pipeline}
\end{figure*}

Our pipeline consists of three stages: (1) speaker diarization, (2) audio clip extraction with weighted sampling, and (3) speaker classification. Figure~\ref{fig:pipeline} illustrates the full workflow.

\subsection{Stage 1: Speaker Diarization}

Speaker diarization is the task of partitioning an audio recording into segments labeled by speaker identity---answering ``who spoke when.''  We use \texttt{pyannote.audio} 3.1 \citep{bredin2023pyannote}, an end-to-end neural diarization framework that jointly performs voice activity detection, speaker embedding extraction, and agglomerative clustering within a single pipeline.  We choose \texttt{pyannote} over cascaded alternatives because it reaches competitive diarization error rates on standard benchmarks \citep{park2022review}, handles overlapping speech---common when officials interrupt or the public speaks over the chair---without a separate overlap-detection module, and is free, open-source, and runs on a single consumer-grade GPU.

Before diarization, each audio file is resampled to 16\,kHz mono via \texttt{librosa} to match the model's expected input format.  After diarization, we discard segments shorter than 0.5\,s, which typically correspond to microphone noise, coughs, or brief crosstalk rather than substantive speech. For each meeting, diarization produces a set of segments $\{(s_i, e_i, k_i)\}_{i=1}^{N}$, where $s_i$ and $e_i$ are the start and end times of segment $i$ in seconds, and $k_i \in \{1, \ldots, K\}$ is the assigned speaker label.

As a transparent baseline for the MLLM, we also extract a 36-dimensional prosodic/spectral feature vector per segment with \texttt{librosa} (pitch, energy, zero-crossing rate, spectral shape, and MFCCs, each summarized by mean and standard deviation; Section~\ref{sec:validation}). We validate diarization against human judgment (Section~\ref{sec:validation}), keeping two quantities distinct: \emph{inter-rater reliability}---agreement among annotators, which is high (mean pairwise Cohen's $\kappa = 0.84$, ``almost perfect'' \citep{landis1977measurement}) and bounds achievable accuracy---and \emph{accuracy}---agreement between the system and the human consensus, 89.4\% across the five validation cities. The residual errors are dominated by a single failure mode (overlapping speech merged into one segment) and are largely absorbed by the downstream error-propagation correction (Appendix~\ref{sec:appendix_error_prop}).

\subsection{Stage 2: Weighted Sampling}

City council meetings are dominated by officials. The mayor, council members, clerk, and staff speak repeatedly throughout a meeting, while most members of the public speak only once, briefly, during public comment. In our ground-truth data, officials account for roughly 83\% of total audio duration. If we sampled audio simply in proportion to how much each person talks, members of the public---the very speakers we most want to characterize---would be severely under-represented, and we would learn little about their participation. We therefore use \emph{weighted sampling}, which represents each speaker roughly equally regardless of how long they hold the floor. Formally, let $T_j = \sum_{i:\,k_i=j}(e_i - s_i)$ denote the total speaking time of speaker~$j$; under naive uniform segment sampling the probability of drawing a segment from speaker~$j$ is proportional to $T_j$, so the heaviest speakers crowd out everyone else. Weighted sampling removes this dependence on $T_j$.

To ensure balanced representation, each unique speaker contributes at most 30 seconds of audio, regardless of total speaking time. If $T_j \leq 30$\,s, all available segments from speaker~$j$ are included. If $T_j > 30$\,s, we select up to three representative clips---the first, middle, and last segment by timestamp---to capture within-speaker variation across the meeting (e.g., a council member may speak formally during roll call but conversationally during discussion). Each selected clip is extracted with 5\,s of audio context before and after the segment boundaries (capped at file start and end) to provide the MLLM with surrounding acoustic context.  This padding allows the model to hear the transition between speakers---for instance, a chair calling on a member of the public before they begin speaking---which can be informative for classification.

\subsection{Stage 3: Speaker Classification}
\label{sec:classification}

We classify each speaker as an \emph{official} (elected representative or government staff) or a \emph{public participant} (member of the public) using Google Gemini 2.5 Flash \citep{gemini2024}, a multimodal large language model (MLLM) capable of processing audio inputs directly via its Files API\@.\footnote{At the time of analysis Gemini 2.5 Flash was among the few production MLLMs accepting raw audio. The classifier is an interchangeable component; we report absolute accuracies for a fixed model and release code so the benchmark can be re-run as models improve.}  All API calls use temperature~$= 0$, so each classification is deterministic given identical input---making results replicable and letting the bootstrapping loop converge to a fixed point rather than drift (Appendix~\ref{sec:appendix_theory}). The exact model identifier, API settings, retry and unparsable-response handling, run dates, token counts, and cost are in Appendix~\ref{sec:appendix_repro}. We develop four classification strategies of increasing sophistication, each exploiting a different source of contextual information.

\subsubsection{Baseline: Zero-Shot and Few-Shot Classification}
\label{sec:baseline}

In the \textbf{zero-shot} setting, the model receives a single audio clip and a text prompt that defines the two categories---\emph{official} (elected council members, city staff, legal counsel, clerks) and \emph{public participant} (residents, community members, and other members of the public)---and instructs the model to return a one-word classification.  The prompt does not provide examples; classification relies entirely on the model's pretraining knowledge of how officials and members of the public sound in public meetings.

The \textbf{few-shot} setting prepends $k$ labeled reference clips before the query.  Each exemplar is presented as an audio file followed by its ground-truth label (e.g., ``This speaker is: official'').  We experimented with 4-shot and 5-shot configurations and varied the class balance and prompt wording (Appendix~\ref{sec:appendix_experiments}).  The best-performing configuration uses $k = 5$ exemplars (4~officials $+$ 1~public participant) drawn randomly from a validated pool, paired with a minimal instruction prompt.  Full prompt text is provided in Appendix~\ref{sec:appendix_experiments} for reproducibility.

\textbf{Response parsing.}  The model's text response is parsed for the tokens ``official'' or ``citizen'' (case-insensitive).\footnote{The deployed prompt and parser use the word \emph{citizen} as the class token. Throughout the paper we report this class as \emph{public participant} (member of the public), since legal citizenship is not observed and the label denotes an institutional role, not legal status.}  If neither token appears, the segment is labeled ``unknown'' and excluded from evaluation.  In practice, $\geq$95\% of responses are parseable under the 5-shot v3 prompt.

\subsubsection{Speaker-Level Aggregation}

Individual clip-level predictions are aggregated to the speaker level using majority voting.  If speaker~$j$ has $n_j$ clips, the final label is the most frequent classification across clips.  This mitigates the impact of noisy individual predictions and diarization merge errors (where two speakers are grouped under one label).

\subsubsection{Temporal Exemplar Selection}
\label{sec:temporal}

Standard few-shot classification selects exemplars from a fixed, meeting-external pool.  We propose an alternative: select exemplars based on \emph{temporal adjacency} within the same meeting.  The problem this solves is that a speaker heard in isolation can be genuinely ambiguous---a brief, informal remark by an official can sound like a member of the public---whereas the same speaker's role is often obvious from \emph{when} they speak relative to others.  The intuition is that council meetings have role-clustered turn-taking: procedural business runs official-only, and public comment produces either an official$\,\rightarrow\,$public$\,\rightarrow\,$official alternation or long runs of consecutive members of the public. Under either format a speaker's nearest-in-time neighbors are disproportionately of the same role, so the temporal neighborhood of a query is highly informative about its role.

This design is not without risk, and the failure mode is instructive. When public participation is very sparse, a member of the public's nearest neighbors may be overwhelmingly officials, biasing the prediction toward ``official''; we observe exactly this in one unusually official-heavy meeting (Appendix~\ref{sec:appendix_tables}). We therefore do not rely on temporal context alone but combine it with meeting-external exemplars and a sliding-window classifier in an ensemble (Section~\ref{sec:ensemble}), and we test directly whether accuracy varies with a speaker's temporal entry position in Section~\ref{sec:robustness}.

Algorithm~\ref{alg:temporal} formalizes the selection procedure.  Given a query speaker~$q$ whose first appearance in meeting~$m$ occurs at time~$t_q$, we rank all other speakers in~$m$ by the absolute difference between their first-appearance times and~$t_q$, then select the~$k$ nearest speakers as exemplars.

We set $k = 4$ throughout.  By drawing exemplars from the same meeting, the model receives audio that shares recording conditions (microphone, room acoustics, background noise) and meeting-specific turn-taking patterns with the query.

\subsubsection{Bootstrapping: Iterative Refinement Without Oracle Labels}
\label{sec:bootstrapping}

Temporal exemplar selection faces a chicken-and-egg problem: to classify a speaker from the labels of their temporal neighbors, we need those neighbors' labels---but those are precisely what we are trying to predict.  We resolve this circular dependency through iterative bootstrapping (Algorithm~\ref{alg:bootstrap}). The idea is to start from an imperfect but usable guess and refine it. In Pass~0, we label every speaker with the meeting-external baseline classifier, which ignores neighbors. In each subsequent pass, we re-classify every speaker, now supplying the \emph{previous} pass's labels for that speaker's temporal neighbors as exemplars. Because the Pass~0 guess is already roughly 90\% accurate, most neighbor labels are correct from the start; each pass therefore feeds slightly cleaner context into the next, the predictions improve, and the labels stabilize once the process reaches a fixed point. We cap the number of passes at $T$ and stop early if no labels change between consecutive passes.

Figure~\ref{fig:bootstrap} illustrates the iterative loop.

This mirrors iterative self-refinement in LLM prompting \citep{sun2024itercot}, where early passes yield the largest gains and the process converges quickly---a pattern we also observe (Section~\ref{sec:validation}).

\subsubsection{Sliding-Window Context Classification}

A third supporting component classifies each speaker from a small window of surrounding context.  The idea is simple: rather than classifying a speaker in isolation, we give the model three consecutive speakers from the meeting and ask it to classify the middle one.  A city council meeting is a temporal sequence of speaker segments $[s_1, s_2, \ldots, s_N]$.  We slide a window of three segments across this sequence, one position at a time, applying the classifier at each position to label the center speaker.  (This is analogous to a one-dimensional convolutional neural network, or 1D CNN, in which a fixed-size filter slides across a sequence and produces a local output at each position; the formal parallel is developed in Appendix~\ref{sec:appendix_theory}.)  Figure~\ref{fig:window} illustrates the procedure.

We evaluate two variants:
\begin{itemize}[nosep]
    \item \textbf{Variant 2a (audio only):} For each query segment~$s_i$, the agent receives the audio of $[s_{i-1},\, s_i,\, s_{i+1}]$ in chronological order and classifies the middle speaker.  No labels are provided for neighbors; the model must infer the meeting pattern directly from hearing consecutive speakers.
    \item \textbf{Variant 2b (audio $+$ labels):} Same as~2a, but the agent also receives the Pass~3 bootstrapped labels for the neighbor segments (e.g., ``Audio~1 was classified as: official'').  This combines raw audio context with prior predictions, giving the model both acoustic and categorical cues.
\end{itemize}

\noindent The overlapping receptive fields mean that adjacent agents share two of their three input segments.  This redundancy enables the detection of turn-taking patterns (e.g., the official$\,\rightarrow\,$public$\,\rightarrow\,$official transition during public comment) at every position in the sequence.

\subsubsection{Ensemble Methods}
\label{sec:ensemble}

The three production strategies---5-shot baseline, Pass~3 bootstrapping, and window~2b---make complementary errors: the 5-shot baseline over-predicts the public class (high recall, lower precision) while the temporal methods are more conservative. We combine them with the majority-vote, AND, and OR rules formalized in the problem statement above. Majority vote is our default; the AND and OR rules expose a precision--recall frontier for applications that weight false positives and false negatives unequally. Performance across all methods and ensembles---evaluated with public-class precision, recall, and F1---is reported in Section~\ref{sec:validation}. Ground truth was built by three independent annotators over eight rounds with majority-vote resolution and Fleiss' $\kappa$ reliability; the protocol and full statistics are in Appendix~\ref{sec:appendix_ground_truth}.\footnote{1,824 consensus-labeled speakers across eight annotation sites (five production cities plus three development sites) and 171 meetings; the $N=1{,}761$ production validation set is the five-city subset with valid predictions from all three classifiers.}


\section{Validation}
\label{sec:validation}

\subsection{Data Partitions and Leakage Controls}
\label{sec:partitions}

Because the classifier requires no task-specific training, the pipeline has no conventional train/test split; but several \emph{operating choices}---the prompt, the fixed exemplar pool, the number of bootstrapping passes, and the ensemble rule---were selected empirically, and honesty about evaluation validity requires stating on which data. Table~\ref{tab:partitions} lists every stage, the data it used, and whether that data overlaps the final evaluation set.

\begin{table*}[t]
\centering
\caption{Data used at each pipeline stage and its overlap with the final evaluation set. The pipeline fits no model weights, but prompt, exemplar, pass, and ensemble choices were selected on labeled data; we report where. The production corpus is held out from every selection stage.}
\label{tab:partitions}
\footnotesize
\renewcommand{\arraystretch}{1.2}
\begin{tabularx}{\textwidth}{@{}p{0.19\textwidth}p{0.30\textwidth}cX@{}}
\toprule
\textbf{Stage} & \textbf{Data used} & \textbf{Overlaps eval set?} & \textbf{Role} \\
\midrule
Prompt development & Development sites (Albany, Chino), 4 meetings, $\approx$51 speakers & No & Tune prompt wording/format and select the 5-shot configuration. \\
Fixed few-shot exemplar pool & A fixed pool of validated exemplar clips (4 officials $+$ 1 public participant per query) & No\textsuperscript{\dag} & Supply meeting-external reference clips to the 5-shot classifier. \\
Temporal exemplars (bootstrapping) & Same-meeting neighbors of each query, labeled by the \emph{previous pass's predictions} & Speakers yes, labels no & In-context neighbors for Pass~1--3; uses predicted, not oracle, labels---so no ground-truth leakage. \\
Pass \& ensemble selection & Validation ground truth ($N=1{,}761$, 5 cities, 8 rounds) & \textbf{Yes} & Choose the converged pass ($T=3$) and the majority-vote rule. \\
Final evaluation & Same validation ground truth ($N=1{,}761$) & --- & Reported accuracies (Tables~\ref{tab:classification},~\ref{tab:convergence}). \\
Production application & 29,238 speakers, 5 cities, unlabeled & Held out & Participation measures (Section~\ref{sec:results}). \\
\bottomrule
\end{tabularx}
\begin{flushleft}
\footnotesize\textsuperscript{\dag}\,\todo{confirm the fixed exemplar clips are disjoint from the 1{,}761 evaluation speakers; if any overlap, exclude them and re-run, or report the overlap count.} \emph{Leakage note:} pass and ensemble selection used the same labeled set as final evaluation (there is no separate held-out test set), so the reported operating-point accuracies are mildly optimistic. We bound this exposure: the selected pass and the runner-up differ by $0.2$ points (Table~\ref{tab:convergence}), and the ensemble improves on the best single method by about $1$ point; both operating choices were fixed by \emph{a priori} rules (convergence and majority) rather than by maximizing validation accuracy (see below). \todo{Action~3: validate a subset of production labels against independent sources---council rosters, agendas, minutes, or speaker cards---and report agreement.}
\end{flushleft}
\end{table*}

\paragraph{Why Pass~3, when Pass~2 scores higher?} Table~\ref{tab:convergence} shows Pass~2 edging Pass~3 on the labeled set (86.9\% vs.\ 86.7\% accuracy; F1 $0.826$ vs.\ $0.824$)---a gap of roughly three speakers, within noise. We nonetheless fix $T=3$, and deliberately \emph{not} by picking the highest-accuracy pass: doing so would select the operating point on the same set used to report it. Instead we stop at the \emph{convergence} point. The per-pass label churn falls $171\!\to\!93\!\to\!83$ and the net change from Pass~2 to Pass~3 is near zero, so Pass~3 is the stable fixed point the bootstrapping procedure is designed to reach (Section~\ref{sec:bootstrapping}). Pass~3 is also not the reported result in isolation---it is one of three members of the majority-vote ensemble, which is what the paper reports. Choosing the converged pass by rule, rather than the argmax pass by hindsight, is the more defensible and less leakage-prone choice, at a cost of $0.2$ accuracy points.

\subsection{Diarization Accuracy}

Table~\ref{tab:diarization} reports diarization validation results across five cities---the Chino development meetings and four production cities (Asbury Park, Brockton, Inglewood, and Jersey City), a total of 483 speaker-pair judgments. The full evaluation protocol and inter-rater reliability analysis are described in Appendix~\ref{sec:appendix_diarization}.

\begin{table*}[t]
\centering
\caption{Diarization Accuracy by City (Majority Vote of 3 Raters)}
\label{tab:diarization}
\begin{tabular}{lcccc}
\toprule
City & Overall & Same-Speaker & Different-Speaker & $N$ Pairs \\
\midrule
Chino, CA        & 87.1\% & 80.2\% & 100.0\% & 124 \\
Asbury Park, NJ  & 85.0\% & 77.5\% & 92.5\%  & 80 \\
Brockton, MA     & 88.8\% & 87.5\% & 90.0\%  & 80 \\
Inglewood, CA    & 91.1\% & 90.0\% & 92.3\%  & 79 \\
Jersey City, NJ  & 94.2\% & 93.3\% & 95.0\%  & 120 \\
\midrule
\textbf{Overall} & \textbf{89.4\%} & \textbf{85.4\%} & \textbf{94.1\%} & \textbf{483} \\
\bottomrule
\end{tabular}
\end{table*}

\paragraph{Standard diarization metrics.} The diarization literature also reports \emph{Diarization Error Rate} (DER) and \emph{Jaccard Error Rate} (JER), which require dense, timestamp-level reference annotations; three research assistants densely annotated seven complete meetings (five double-annotated) and we scored pyannote with \texttt{pyannote.metrics}. Pyannote achieves a speech-weighted DER of 7.6\% and JER of 21.9\% (Appendix~\ref{sec:appendix_der}, Table~\ref{tab:der}), consistent with the 89.4\% pairwise agreement and competitive with reported benchmark accuracy. Errors concentrate almost entirely in Brockton's noisier subcommittee audio (DER 17--22\%; other cities $\leq$6.8\%), yet Brockton still attains among the highest role-classification accuracy, because the clip-level majority vote absorbs the diarization splits and merges that inflate JER rather than propagating them to speaker-role labels.

\subsection{MLLM Classification Accuracy}

We evaluate the three production methods---5-shot (fixed exemplars), Pass~3 bootstrapping, and window~2b (audio + labels)---and their ensembles on the $N = 1{,}761$ ground-truth speakers (Section~\ref{sec:ground_truth_overview}); all runs use Gemini~2.5~Flash at temperature~0. Table~\ref{tab:classification} reports the full results. Among individual methods, Pass~3 is best (86.7\% accuracy, public F1 $= 0.824$), consistent with temporal context adding signal, with window~2b (85.2\%) and the 5-shot baseline (84.7\%) close behind. Because the three make complementary errors---the 5-shot baseline favors precision (0.878/0.710) while Pass~3 balances precision and recall (0.859/0.792)---their \textbf{majority vote} is the best overall method: \textbf{87.7\% accuracy, $\kappa = 0.736$, public F1 $= 0.832$}. These pairwise differences are statistically significant: by McNemar's test (Appendix~\ref{sec:appendix_validation}), the ensemble improves on the 5-shot baseline ($p < 0.001$) and on window~2b ($p < 0.001$), and Pass~3 improves on 5-shot ($p = 0.015$); the ensemble's edge over Pass~3 alone is not significant ($p = 0.097$), consistent with their overlapping errors. The AND and OR rules trace the precision--recall frontier for applications that weight the two errors unequally: AND (5-shot~$\cap$~window~2b) gives the highest public precision (0.930), OR (Pass~3~$\cup$~window~2b) the highest recall (0.854), each at a cost on the other axis (theory in Appendix~\ref{sec:appendix_theory}).

\begin{table*}[t]
\centering
\caption{Classification Performance on Production Ground Truth (Gemini 2.5 Flash, $N = 1{,}761$ speakers, 5 cities, 8 annotation rounds)}
\label{tab:classification}
\begin{tabular}{lcccccc}
\toprule
Method & $N$ & Accuracy & $\kappa$ & Pub.\ P & Pub.\ R & Pub.\ F1 \\
\midrule
\multicolumn{7}{l}{\textit{Individual methods}} \\
5-shot (fixed exemplars)         & 1,761 & 0.847 & 0.669 & 0.878 & 0.710 & 0.785 \\
Pass 3 (bootstrapped)            & 1,761 & 0.867 & 0.718 & 0.859 & 0.792 & 0.824 \\
Window 2b (audio + labels)       & 1,761 & 0.852 & 0.685 & 0.844 & 0.766 & 0.803 \\
\midrule
\multicolumn{7}{l}{\textit{AND ensembles}} \\
AND: 5-shot $\cap$ Pass 3        & 1,761 & 0.848 & 0.665 & 0.924 & 0.668 & 0.775 \\
AND: 5-shot $\cap$ Window 2b     & 1,761 & 0.844 & 0.654 & \textbf{0.930} & 0.652 & 0.766 \\
AND: Pass 3 $\cap$ Window 2b     & 1,761 & 0.859 & 0.692 & 0.919 & 0.704 & 0.797 \\
\midrule
\multicolumn{7}{l}{\textit{OR ensembles}} \\
OR: 5-shot $\cup$ Pass 3         & 1,761 & 0.867 & 0.721 & 0.828 & 0.834 & 0.831 \\
OR: 5-shot $\cup$ Window 2b      & 1,761 & 0.856 & 0.698 & 0.812 & 0.824 & 0.818 \\
OR: Pass 3 $\cup$ Window 2b      & 1,761 & 0.860 & 0.710 & 0.803 & \textbf{0.854} & 0.828 \\
\midrule
\multicolumn{7}{l}{\textit{Majority vote}} \\
Majority (2/3)                   & 1,761 & \textbf{0.877} & \textbf{0.736} & 0.895 & 0.777 & \textbf{0.832} \\
\bottomrule
\end{tabular}
\begin{flushleft}
\footnotesize\textit{Note:} Meeting-clustered bootstrap 95\% CIs on accuracy (3{,}000 resamples of meetings, since speakers within a meeting are correlated) for the headline rows: 5-shot $[82.8, 86.6]$, Pass~3 $[84.8, 88.6]$, window~2b $[83.3, 87.2]$, majority vote $[85.8, 89.4]$. These cluster-robust marginal intervals overlap; the \emph{paired} McNemar tests (Appendix~\ref{sec:appendix_validation}) are the appropriate test for the differences and confirm the ensemble improves on 5-shot and window~2b ($p<0.001$).
\end{flushleft}
\end{table*}

\paragraph{Convergence and error structure.} Bootstrapping converges rapidly: accuracy rises from 84.8\% (Pass~0) to a plateau near 86.9\% by Pass~2, with the largest gain in the first iteration and per-pass label churn falling $171\!\to\!93\!\to\!83$ (Appendix~\ref{sec:appendix_validation}, Table~\ref{tab:convergence}). The corrections are net-positive and disproportionately rescue true members of the public misclassified as officials (127 of 190 fixes), unwinding the zero-shot majority-class bias; because the net change from Pass~2 to Pass~3 is near zero, we fix $T=3$ at that fixed point rather than iterating further. Window~2b's errors are partially orthogonal to those of the other two methods, which is what makes it a useful third ensemble member despite its lower solo accuracy.

Performance varies across cities: the majority-vote ensemble achieves 90.8\% accuracy in Brockton ($N = 283$) and 89.0\% in Inglewood ($N = 318$), with public F1 ranging from 0.711 in Brockton (where the low public-participant rate of 17.3\% in the ground truth sample makes classification harder) to 0.873 in Jersey~City ($N = 534$). These per-city differences underscore the value of evaluating on geographically diverse ground truth rather than a single development site.

\subsection{Comparison with Supervised Audio Baselines}

A natural question is whether the MLLM is doing real work, or whether much simpler audio methods would classify speakers just as well. To answer it we ask: how well can a conventional, purpose-trained classifier do on the same task, using only measurable properties of the audio signal? This matters because the simpler methods are cheaper and fully transparent; if they matched the MLLM, the added complexity and opacity of a large model would be hard to justify. Table~\ref{tab:prosodic} reports the comparison; we summarize the baselines here and defer the full feature list to Appendix~\ref{sec:appendix_validation}.

\paragraph{Held-out test set.} Because the operating points are fixed by \emph{a priori} rules---$T=3$ by convergence and a majority-vote ensemble---rather than by maximizing accuracy on the labeled set, the labeled speakers serve as both a selection and a reporting set (a limitation we flagged above). To show this induces no selection bias, we evaluate on data never used for any pipeline choice. On a difficulty-matched random split by meeting (50/50, seed~42), the ensemble scores \textbf{87.4\%} on the held-out half versus 87.8\% on the selection half---statistically indistinguishable, confirming that the \emph{a priori} rules do not overfit. As a stricter, temporally held-out check, we designate the later annotation rounds (R9--R11; $N=681$ speakers, 65 meetings; never used to choose exemplars, the number of passes, or the ensemble rule) as a test set: there the ensemble reaches \textbf{85.0\%}, lower only because these later rounds are more public-comment-heavy and therefore intrinsically harder, yet still well above the strongest supervised baseline (80.3\%) and the majority-class rate (60.7\%). Both splits are produced by \texttt{experiments/heldout\_test.py}.

\paragraph{Alternative input modalities and models.} A prior question is whether the input \emph{modality}---audio---is doing the work, or whether transcripts, video, or an open model would classify speakers just as well. Table~\ref{tab:modality} compares the audio-native pipeline against three single-pass alternatives, each stripped of our temporal refinements to isolate the effect of input modality and model. Three findings stand out. First, transcription is not lossless: an ASR-plus-text pipeline (Whisper-large-v3 $\rightarrow$ Gemini) reaches 81.8\%, below the audio-native model on the same speakers (85.7\%), and degrades further on noisy public-comment audio (Appendix~\ref{sec:appendix_audio_vs_text}). Second, video alone is weaker still: 75.7\% accuracy, and only for the 80\% of speakers the camera actually frames (it holds a fixed dais shot for the rest; the on-screen rate ranges 45--100\% by city). Third, the leading open audio-LLM (Qwen2-Audio-7B) collapses to a single class and provides no discriminative signal---despite transcribing the audio accurately---indicating that audio-native role classification is not yet reproducible with open models, even as our closed-model pipeline reaches 87.7\%. Our temporal refinements add a further $+3$ points over the audio-native base (84.7\%$\rightarrow$87.7\%).

\begin{table*}[t]
\centering
\caption{Input-modality and model comparison. Each alternative is single-pass (no bootstrapping or sliding-window refinement) to isolate the effect of input modality and model; our audio-native rows then show the gain from temporal structure. Sample sizes differ by data availability (see notes). Majority-class baseline: 60.7\%.}
\label{tab:modality}
\begin{tabular}{lllcc}
\toprule
Method & Input & Model & $N$ & Acc. \\
\midrule
\multicolumn{5}{l}{\textit{Alternative modalities / models (single-pass)}} \\
Open audio-LLM             & audio        & Qwen2-Audio-7B (open)         & 599     & ---$^{a}$ \\
Video-only MLLM            & video frames & Gemini 2.5 Flash              & 381     & 0.757$^{b}$ \\
ASR $\rightarrow$ text LLM & transcript   & Whisper-lg-v3 + Gemini        & 593     & 0.818 \\
\midrule
\multicolumn{5}{l}{\textit{Audio-native MLLM (ours)}} \\
Base (5-shot)                  & audio & Gemini 2.5 Flash & 1{,}761 & 0.847 \\
\quad + bootstrapping (Pass 3) & audio & Gemini 2.5 Flash & 1{,}761 & 0.867 \\
\quad + ensemble (majority)    & audio & Gemini 2.5 Flash & 1{,}761 & \textbf{0.877} \\
\bottomrule
\end{tabular}
\begin{flushleft}
\footnotesize
$^{a}$ Qwen2-Audio-7B \emph{collapses to a single class}, giving no discriminative signal: 39.9\% when it defaults to the public-participant class (below the majority baseline) or 61.1\% when official-heavy prompting flips it to all ``official''---despite transcribing the audio accurately. \\
$^{b}$ Among the 80\% of speakers the camera frames; video-only cannot decide for the rest. \\
Subsets differ by data availability (video needs the speaker on-screen; ASR / open-audio need locally available source audio). On the shared 593-speaker ASR subset: single-pass audio scores 0.857 vs.\ 0.818 for single-pass ASR+text; giving \emph{text} the same temporal machinery (bootstrapping Pass~3 + window~2b ensemble) closes neither the modality gap nor the absolute level---text ensemble reaches 0.826 vs.\ audio ensemble 0.895---confirming the advantage is a genuine modality effect rather than a pipeline asymmetry (\texttt{experiments/w4\_text\_temporal\_ensemble.py}).
\end{flushleft}
\end{table*}

The supervised baselines require hand-crafted acoustic features and pretrained speaker embeddings, which were extracted for the subset of the validation speakers drawn from annotation rounds R2 and R4--R7 ($N = 1{,}081$; 653 officials, 428 public participants). We therefore report this comparison on that common $N = 1{,}081$ subset, on which the MLLM majority-vote ensemble scores 89.3\% (versus 87.7\% on the full $N = 1{,}761$ set); the two rates differ only because the subset omits the later, more public-comment-heavy rounds, and the ranking of methods is unchanged. Table~\ref{tab:prosodic} reports the comparison on these $N = 1{,}081$ speakers.  Two families of supervised methods are evaluated: prosodic feature classifiers (79 hand-crafted acoustic and temporal features aggregated from diarization output) and pretrained speaker embedding classifiers (ECAPA-TDNN and Wav2Vec2).  The best supervised model is the prosodic Gradient Boosting classifier at 80.3\% accuracy ($\kappa = 0.580$, public F1 $= 0.736$), followed by ECAPA-TDNN + Random Forest at 77.4\% ($\kappa = 0.510$).  All supervised baselines fall substantially below the zero-shot MLLM 5-shot method (85.3\%) and the majority-vote ensemble (89.3\%).

\section{Application}
\label{sec:results}

We apply the full pipeline to 1,565 meetings across five cities (Asbury Park, Brockton, Inglewood, Jersey City, and New Brunswick), classifying 29,238 unique speakers.\footnote{East Palo Alto is excluded from substantive analyses due to its small corpus (2 meetings). Classification uses the majority-vote ensemble of Pass~3, 5-shot, and window~2b labels for each speaker.} Table~\ref{tab:results_summary} summarizes the cross-city results.

\begin{table*}[t]
\centering
\caption{Cross-City Summary of Speaker Classification Results}
\label{tab:results_summary}
\begin{tabular}{lrrrrrrr}
\toprule
 & & & Official & Public & \multicolumn{2}{c}{Public Floor Share (\%)} & Gini \\
\cmidrule(lr){6-7}
City & Meetings & Speakers & (\%) & (\%) & Pooled & Median & \\
\midrule
Asbury Park   & 364 & 7,270 & 70.7 & 29.3 & 19.2 & 16.3 & 0.541 \\
Brockton      & 499 & 6,600 & 86.6 & 13.4 &  6.0 &  0.3 & 0.531 \\
Inglewood     & 141 & 2,346 & 65.1 & 34.9 & 20.6 & 17.0 & 0.456 \\
Jersey City   & 299 & 9,143 & 58.6 & 41.4 & 25.2 & 17.1 & 0.548 \\
New Brunswick & 262 & 3,879 & 71.4 & 28.6 & 27.7 & 20.5 & 0.529 \\
\midrule
\textbf{All}  & \textbf{1,565} & \textbf{29,238} & \textbf{70.1} & \textbf{29.9} & \textbf{18.7} & \textbf{10.0} & \textbf{0.533} \\
\bottomrule
\end{tabular}
\begin{flushleft}
\footnotesize\textit{Note:} Public (\%) is the pooled share of detected speakers classified as members of the public. \emph{Public Floor Share} is the share of total meeting speaking time held by the public, reported two ways: \emph{Pooled} aggregates speaking time over the city's entire corpus (comparable, on the same basis, to the pooled headcount share), while \emph{Median} is the median across meetings (the typical meeting). The two diverge because the meeting-level distribution is right-skewed and heavy at zero---25.8\% of meetings corpus-wide have no public speaker. Gini is the median within-meeting Gini coefficient of speaking time across all speakers.
\end{flushleft}
\end{table*}

\begin{figure}[t]
\centering
\includegraphics[width=\columnwidth]{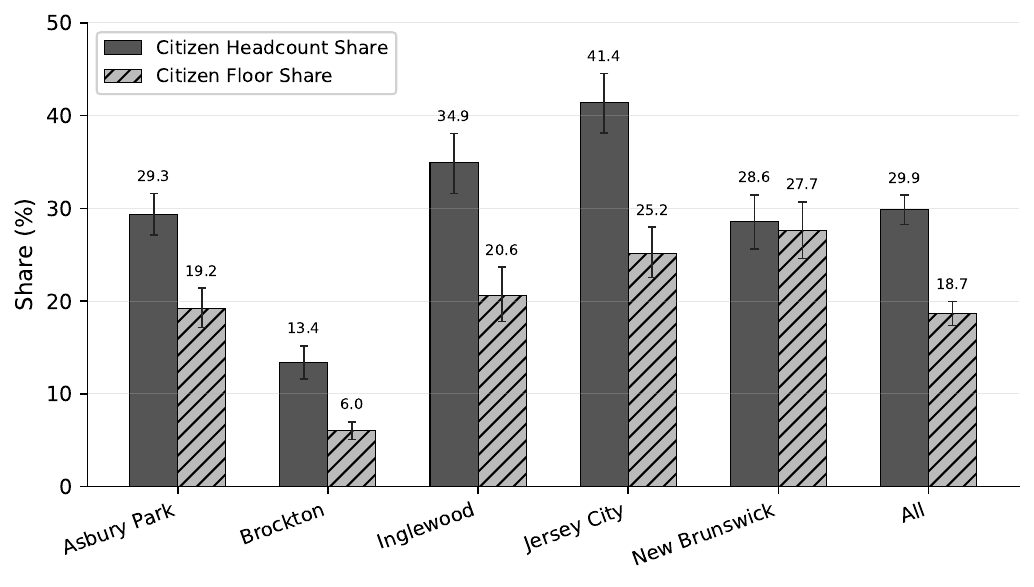}
\caption{The Participation Gap: public headcount share vs.\ public floor share by city, both computed as pooled (corpus) shares so the two are on the same basis and directly comparable. In every city, the public's share of speaking time falls below its share of speakers. Pooled across all five cities, members of the public are 29.9\% of speakers but hold 18.7\% of the floor; Table~\ref{tab:results_summary} additionally reports the meeting-median floor share (10.0\% overall). Error bars are 95\% meeting-clustered bootstrap confidence intervals; bias-corrected floor shares are reported in Appendix~\ref{sec:appendix_floorshare_correction}. ``Showing up'' does not translate into proportionate voice.}
\label{fig:participation_gap}
\end{figure}

Overall, 70.1\% of detected speakers are classified as officials and 29.9\% as members of the public (these are observed, plug-in proportions; bias-corrected estimates that account for classification error are reported in Appendix~\ref{sec:appendix_error_prop} and only \emph{amplify} the cross-group differences described below). But this headcount understates how concentrated the floor is. Measured on the same (pooled) basis, members of the public are 29.9\% of speakers yet receive only 18.7\% of total speaking time---a floor-to-headcount ratio of 0.63 (95\% CI $[0.61, 0.65]$; the 11.2-point gap has CI $[10.4, 11.9]$, excluding zero); the shortfall is comparable within the typical meeting, where the median public headcount share (17.6\%) exceeds the median public floor share (10.0\%). We report the pooled and meeting-median quantities separately, and avoid pairing statistics computed on different bases, because the meeting-level distribution is highly skewed: 25.8\% of meetings have no public speaker at all, and per-meeting public floor share spans an interquartile range of 0.0\%--24.8\% (mean 15.6\%, median 10.0\%). Both gaps are robust to classification error: a duration-weighted misclassification correction raises the corrected pooled public floor share to 22.2\% (95\% CI $[19.1, 26.3]$), still well below the corrected public headcount share of 33.4\% (Appendix~\ref{sec:appendix_floorshare_correction}), so the shortfall is not an artifact of the classifier. Figure~\ref{fig:participation_gap} visualizes this participation gap across all five cities. The concentration of speaking time is pronounced: the top speaker is an official in 94.2\% of meetings, and the top three speakers are all officials in 76.5\% of meetings. Members of the public also enter meetings later---at the median 48\% mark versus 25\% for officials---consistent with standard council formats in which procedural business precedes public comment. This entry-timing regularity is genuine (it holds when the feature-importance model is refit on human labels) but does not drive the classifier: entry time alone classifies speakers at only 67.4\% accuracy and a temporal-features-only model at 75.5\%, well below the ensemble's 87.7\%, so the instrument's accuracy rests on acoustic and lexical content rather than on position (Appendix~\ref{sec:appendix_tables}, Table~\ref{tab:temporal_ablation}). These patterns extend \citeauthor{einstein2019still}'s (\citeyear{einstein2019still}) finding about \emph{who shows up} to local meetings: our continuous measure captures \emph{how much room they get}, and the answer is substantially less than headcounts alone would suggest. We do not treat headcount parity as a normative benchmark: officials chair the meeting and conduct its procedural business, and public comment is typically capped at a few minutes per speaker, so part of this gap is structural by design. The value of the floor-time measure is therefore descriptive---it quantifies how much room the public receives under each institution's rules---and the substantive signal is the cross-jurisdiction \emph{variation} in that share (Table~\ref{tab:results_summary}), not any single-meeting deviation from equality.

Cross-city variation illustrates the pipeline's ability to detect institutional differences, which we report descriptively rather than as causal estimates. The median public floor share ranges from 20.5\% in New Brunswick to just 0.3\% in Brockton---a nearly 70-fold difference (on a pooled basis the same contrast is 27.7\% vs.\ 6.0\%; Table~\ref{tab:results_summary}). Brockton's near-zero figure warrants a direct look, and coding each meeting's type from its title clarifies it (Appendix~\ref{sec:appendix_tables}, Table~\ref{tab:meeting_type}). Only 16.8\% of Brockton's meetings are regular city-council sessions; the rest are school-committee (28.3\%), other committee (22.4\%), board/commission (16.4\%), and subcommittee (14.2\%) meetings that conduct administrative business with little or no open public-comment period. Composition alone, however, does not explain the figure: even Brockton's \emph{regular council} meetings have a median public floor share of just 0.1\% (5.0\% pooled), against 18--37\% for regular council meetings in the other four cities. Brockton is thus a genuine outlier rather than a measurement artifact---its classification accuracy is in line with the other cities (Table~\ref{tab:classification}) and its public \emph{headcount} share (13.4\%) is low but not vanishing---but the pattern reflects both an administrative-heavy corpus and a council format that surfaces almost no public comment on the recording. We flag it as worth confirming against meeting agendas, since formats that route public comment to call-in lines, written submission, or separate hearings could depress the measured share. Beyond Brockton, Jersey City has the highest public headcount share (41.4\%), yet the median public floor share is only 17.1\% (25.2\% pooled); fully 25.8\% of meetings have zero public speakers, and 40.9\% allocate less than 5\% of floor time to the public. These differences are consistent with institutional design---council size, the placement and duration of public comment periods, and meeting format---and the pipeline's ability to detect them at scale opens a path toward systematic comparative research on local democratic institutions.

\begin{figure}[t]
\centering
\includegraphics[width=\columnwidth]{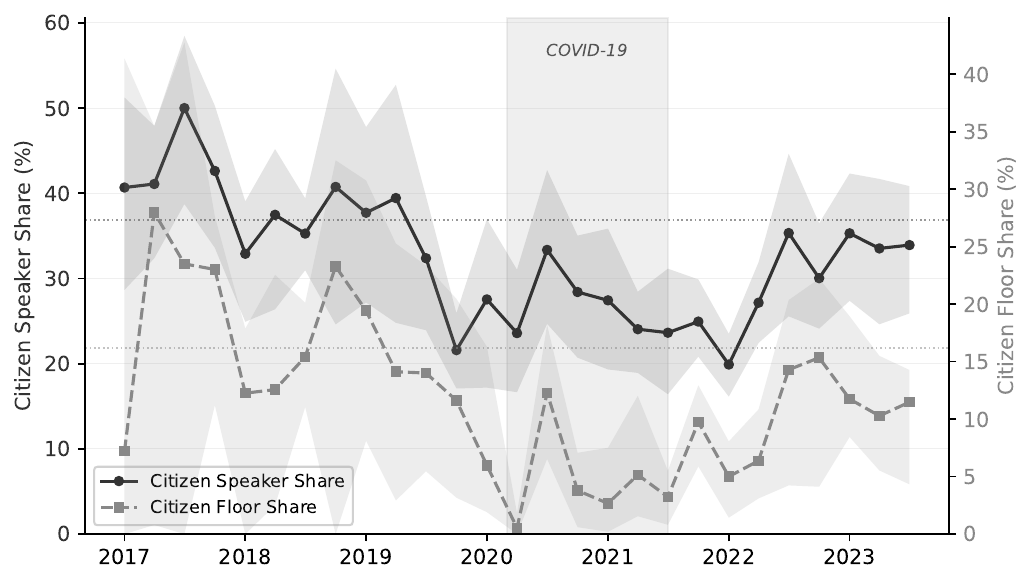}
\caption{Public participation over the COVID-19 period: quarterly public speaker share and floor share, 2017--2023. The shaded vertical region marks the pandemic period (March 2020 -- June 2021); dotted lines indicate pre-COVID baselines. Shaded bands around each series are 95\% meeting-clustered bootstrap confidence intervals. Public participation dropped sharply during the pandemic and had not fully recovered by 2023.}
\label{fig:covid}
\end{figure}

The COVID-19 period illustrates the pipeline's capacity for longitudinal measurement. Public speaker share dropped from 27.4\% pre-pandemic (2017--2019) to 12.5\% during 2020--2021, with only partial recovery to 18.2\% afterward (2022--2023). The decline in floor share (meeting medians) was even steeper: from 15.5\% to 4.2\%, with recovery to just 9.6\% (Figure~\ref{fig:covid}). The decline is not an artifact of classification error: on a pooled (time-weighted) basis, and using period-specific error rates that we estimate rather than assume, the corrected public floor share falls from 38.0\% (95\% CI $[26.7, 52.3]$) pre-pandemic to 16.7\% ($[12.7, 22.4]$) during the pandemic---non-overlapping intervals (Appendix~\ref{sec:appendix_transport}). These patterns are consistent with \citeauthor{einstein2022still}'s (\citeyear{einstein2022still}) evidence that virtual meetings did not diversify participation. We describe them as longitudinal patterns rather than causal estimates: the recovery was incomplete as of 2023, but our design does not identify the pandemic as their cause. Per-city and per-year coverage is also uneven---Brockton, for instance, contributes few pre-2020 recordings---so the aggregate trajectory reflects the available convenience panel rather than a balanced one, and should be read as a pattern in this corpus rather than a representative national estimate. \citet{martin2026participation}, exploiting the staggered withdrawal of remote-participation options across California cities, provide complementary quasi-experimental evidence that meeting access costs shape the volume of public participation.
\todo{Action~4: add meeting-clustered, bias-corrected uncertainty intervals (error bars/bands) to Figures~\ref{fig:participation_gap} and~\ref{fig:covid}, and stratify the classification error rates by period to test the transportability assumption (Appendix~\ref{sec:appendix_transport}).}

\section{Discussion and Conclusion}
\label{sec:discussion}

City council meetings are a primary channel through which the public engages with local government, yet systematic measurement of who speaks---and for how long---has been limited by the cost of manual transcription and coding. This paper develops a measurement instrument showing that audio analysis alone can reliably distinguish officials from members of the public at scale---an audio-as-data approach to democratic participation. The instrument combines neural speaker diarization with multimodal LLM classification, using the temporal structure of council meetings through iterative refinement and a sliding-window classifier as supporting components. A majority-vote ensemble of three complementary classifiers achieves 87.7\% accuracy and a public-participant F1 of 0.832 on human-validated ground truth from five production cities ($N = 1{,}761$ speakers across eight annotation rounds). Applied to 1,565 meetings, the instrument produces the participation measures analyzed in Section~\ref{sec:results}: large gaps between public headcount and floor time, substantial cross-city variation, and a decline in public participation over the COVID period---measures that would be prohibitively expensive to obtain through manual coding.

These results extend \citeauthor{einstein2019still}'s (\citeyear{einstein2019still}) observation that local meetings are dominated by a narrow, unrepresentative slice of the public along a new dimension. Where prior work measured \emph{who shows up} (demographics of speakers drawn from sign-in sheets or manual coding), our measure captures \emph{how much room they get}: the actual allocation of speaking time between government and governed. The substantive payoff is that this allocation is now a measurable outcome that can be linked to its institutional context. The near-70-fold spread in public floor share across our five cities (Section~\ref{sec:results}) is large and co-varies with institutional design features that structure voice---whether public comment is a fixed agenda item or discretionary, the per-speaker time limit, the total time reserved for comment, council size, and whether remote or call-in participation is permitted---rather than with measurement error. Because the instrument observes institutional role only, we make no claim about the demographic composition of who speaks; that would require external data we do not use (Section~\ref{sec:ethics}). By measuring speaking-time shares at scale, researchers can move from documenting that participation is unequal toward explaining \emph{why}: distinguishing jurisdictions that merely permit public speech from those that structurally prioritize it, and---with an appropriate causal design---estimating which institutional rules expand public voice. Testing these relationships requires coding the relevant rules across many jurisdictions and is the goal of ongoing work; the contribution here is the measurement instrument that makes such an analysis possible.

Methodologically, this paper advances an ``Audio-as-Data'' paradigm for political science \citep{mestre2025potential}. The dominant computational approach to studying political speech---Text-as-Data \citep{grimmer2013text}---requires transcription as an intermediate step, inheriting the cost, latency, and systematic bias of automatic speech recognition \citep{koenecke2020racial}. Prior audio-as-data work has focused on elite speech: vocal pitch in legislatures \citep{dietrich2019pitch, rask2025partisan}, emotional arousal at the Supreme Court \citep{dietrich2019emotional, knox2021dynamic}, and campaign advertisement coding \citep{tarr2023automated}. We extend this paradigm to mass democratic participation, using audio-native multimodal LLMs rather than hand-engineered acoustic features, and show that for the specific task of classifying speaker roles transcription is not required: a multimodal LLM operating directly on audio reaches 88\% accuracy---comparable to transcription-based pipelines \citep{martin2026participation} but without their cost or transcription bias---and sufficient for reliable aggregate measurement.

Three methodological insights emerge. First, \emph{meeting structure is an informative signal}: the largest accuracy gains come from exploiting temporal context rather than better prompts or more exemplars. Second, \emph{iterative self-refinement works}: bootstrapping converges within three passes and is net-corrective on the labeled set (190 labels fixed against 157 broken), predominantly by rescuing members of the public misclassified as officials---unwinding the zero-shot majority-class bias. Third, \emph{the MLLM generalizes across cities without retraining}: leave-one-city-out evaluation shows that the majority-vote ensemble maintains 89.5\% macro-average accuracy on held-out cities, while supervised prosodic classifiers drop from 80.3\% to 70.0\% (Section~\ref{sec:robustness}).  This zero-shot generalizability is what makes scaling to new jurisdictions practical. It is also why our baselines are training-free rather than fine-tuned: a model fine-tuned on one set of cities optimizes in-domain fit at the expense of the cross-jurisdiction transfer that makes the instrument deployable at scale---precisely the out-of-domain degradation the supervised baselines already exhibit. Fine-tuning would also reintroduce the per-setting labeling cost the instrument is designed to eliminate.

\subsection{Limitations}
\label{sec:limitations}
Four caveats bound these claims. First, on \emph{scope}: as discussed in Section~\ref{sec:data}, the five cities are urban, suburban, and Democratic-leaning, and the pipeline has been validated only across jurisdictions of this broad type. The substantive patterns we report should be read as proof of concept, not as a national portrait, and transfer to rural or partisan-charged settings remains to be shown. Relatedly, the instrument yields a single binary label---official vs.\ member of the public---and observes neither speaker demographics nor identities; speakers are unique within a meeting, so repeat attendance across meetings is not tracked. Questions about \emph{which} groups dominate, or whether the same individuals return, therefore fall outside what the measure can answer on its own and would require linkage to external data that we deliberately do not perform (Section~\ref{sec:ethics}). Second, on \emph{mechanism}: although operating on raw audio in principle preserves prosodic and turn-taking information that text discards, our classifier does not explicitly extract or analyze these cues, and we do not claim that prosody per se drives its accuracy. In the formal, turn-regulated setting of a council meeting, role is plausibly carried as much by procedural register, self-introduction, and meeting position as by tone of voice; treating paralinguistic signal as a substantive object of study---rather than an implicit input to a classifier---is a direction for future work, not a result we establish here. Third, on \emph{opacity and durability}: the MLLM is a closed model whose behavior may shift across versions. We mitigate this by validating against human ground truth and propagating error formally (Appendix~\ref{sec:appendix_error_prop}), and by releasing code so results can be re-benchmarked, but absolute accuracies are tied to a specific model and should be re-verified as models change. Fourth, on \emph{the value of operating on audio directly}: a head-to-head comparison against the conventional audio-to-text-to-data pipeline (Appendix~\ref{sec:appendix_audio_vs_text}) shows that transcription is not a lossless intermediate step---the audio-direct classifier matches a Whisper-plus-text-classifier baseline on clean speech but is substantially more robust on noisy audio, where ASR degrades. That comparison is currently run on a subset of speakers with a mid-sized ASR model; a definitive version on the full ground truth with a larger ASR model is left for the final draft.

Several extensions are natural. The most immediate is scaling beyond the current sample: the LocalView database \citep{barari2023localview} covers over 2,800 local governments with publicly available recordings, and applying our pipeline to this corpus would produce a nationally representative dataset of public speaking time. A second direction, and the one that turns the instrument into substantive theory, is to code institutional rules---public-comment format, per-speaker and total time limits, council size, remote-participation options---across many jurisdictions and, with a suitable causal design, estimate which features expand public floor share, exploiting the large within-sample variation documented above. A third direction is combining audio classification with selective transcription---using role labels to identify public-comment speech segments for targeted content analysis, reducing transcription costs by an order of magnitude. Fourth, finer-grained speaker taxonomies (elected official, staff, organized interest, individual member of the public) would increase informativeness; end-to-end multimodal architectures such as SpeakerLM \citep{yin2026speakerlm} may make such multi-class classification feasible. Finally, the pipeline generalizes in principle to any deliberative setting with structured turn-taking---school boards, planning commissions, public utility hearings, state legislative committees---wherever institutional insiders and public participants are distinguished by procedural role; the core architecture (diarize $\rightarrow$ sample $\rightarrow$ classify) is setting-agnostic and only the classification prompt requires adaptation, though, as noted, transfer to each new setting should be validated rather than assumed.

\section{Ethical Considerations}
\label{sec:ethics}

\paragraph{Data and human-subjects status.} All audio analyzed in this study consists of public records of open government meetings, published by the cities themselves on their official public YouTube channels. Speakers address a public body on the record and are aware that the proceedings are recorded and posted. The research analyzes these already-public recordings and does not intervene with, interact with, or collect private information from individuals; it was accordingly determined to be \emph{not human-subjects research} (IRB-exempt). The same determination covers the research-assistant annotation task, which labels the institutional role of speakers in public recordings rather than studying the annotators.

\paragraph{De-identification and voice data.} Our analysis assigns de-identified speaker labels (SPEAKER\_00, SPEAKER\_01, \ldots) and classifies each speaker only by institutional role (official vs.\ member of the public), not by individual identity. We do not name, re-identify, or link speakers to voter files, property records, or any external database. Audio clips were transmitted to the Gemini Files API for classification and were not assembled into or released as a derived voice database. \todo{state the third-party (Google) retention terms for audio uploaded to the Files API, and how long clips persist / access controls.}

\paragraph{Demographic and accent-related risks.} Automated speech and audio models can perform unevenly across accents, dialects, and recording conditions. We mitigate this by validating on geographically diverse ground truth, reporting per-city performance, and formally propagating classification error into the substantive estimates (Appendix~\ref{sec:appendix_error_prop}); nonetheless, transfer to populations with different speech patterns should be validated rather than assumed. The instrument classifies institutional \emph{role}, not demographic attributes, and should not be used to infer the race, gender, age, or other characteristics of individual speakers.

\paragraph{Redistribution and intended use.} We release derived, aggregate participation measures, code, and releasable annotations; we do not redistribute the raw meeting audio, which remains available from the cities' public channels. The intended use is aggregate measurement of participation and comparison of institutions. Using the pipeline to identify, profile, surveil, or re-contact individual speakers is outside its intended use and is not supported by its validation.

\section{Generative AI Usage}
\label{sec:genai}

Generative AI is integral to the method itself: the measurement instrument classifies speaker roles with Google \texttt{gemini-2.5-flash}, an audio-native multimodal LLM, as described in Sections~\ref{sec:method}--\ref{sec:validation} and Appendix~\ref{sec:appendix_repro}. Beyond this core methodological use, the authors used \textbf{Claude Code} to assist with software development (writing and refactoring the pipeline and analysis scripts) and to proofread the manuscript. No generative AI system was used to generate, fabricate, or alter research data, results, tables, or substantive scientific claims: all reported measurements were produced by the authors' code and validated against human-annotated ground truth, and all analytical and interpretive claims are the authors' own.

\clearpage

\clearpage

\bibliographystyle{ACM-Reference-Format}
\bibliography{references}

@book{verba1995voice,
  title={Voice and Equality: Civic Voluntarism in {American} Politics},
  author={Verba, Sidney and Schlozman, Kay Lehman and Brady, Henry E.},
  year={1995},
  publisher={Harvard University Press},
  address={Cambridge, MA}
}

@article{brady1995beyond,
  title={Beyond {SES}: A Resource Model of Political Participation},
  author={Brady, Henry E. and Verba, Sidney and Schlozman, Kay Lehman},
  journal={American Political Science Review},
  volume={89},
  number={2},
  pages={271--294},
  year={1995}
}

@article{einstein2019still,
  title={Who Participates in Local Government? {Evidence} from Meeting Minutes},
  author={Einstein, Katherine Levine and Palmer, Maxwell and Glick, David M.},
  journal={Perspectives on Politics},
  volume={17},
  number={1},
  pages={28--46},
  year={2019}
}

@article{einstein2022still,
  title={Still Muted: The Limited Participatory Democracy of {Zoom} Public Meetings},
  author={Einstein, Katherine Levine and Glick, David M. and Palmer, Maxwell and Pei, Luisa},
  journal={Urban Affairs Review},
  volume={58},
  number={6},
  pages={1595--1622},
  year={2022}
}

@article{sahn2024public,
  title={Public Comment and Public Policy},
  author={Sahn, Alexander},
  journal={American Journal of Political Science},
  year={2024},
  note={Forthcoming}
}

@book{einstein2020neighborhood,
  title={Neighborhood Defenders: Participatory Politics and {America's} Housing Crisis},
  author={Einstein, Katherine Levine and Glick, David M. and Palmer, Maxwell},
  year={2020},
  publisher={Cambridge University Press},
  address={Cambridge}
}

@article{barari2023localview,
  title={{LocalView}, A Database of Public Meetings for the Study of Local Politics and Policy-Making in the {United States}},
  author={Barari, Soubhik and Simko, Tyler},
  journal={Scientific Data},
  volume={10},
  number={1},
  pages={135},
  year={2023}
}

@article{koenecke2020racial,
  title={Racial Disparities in Automated Speech Recognition},
  author={Koenecke, Allison and Nam, Andrew and Lake, Emily and Nudell, Joe and Quartey, Minnie and Mengesha, Zion and Toups, Connor and Rickford, John R. and Jurafsky, Dan and Goel, Sharad},
  journal={Proceedings of the National Academy of Sciences},
  volume={117},
  number={14},
  pages={7684--7689},
  year={2020}
}

@inproceedings{radford2023whisper,
  title={Robust Speech Recognition via Large-Scale Weak Supervision},
  author={Radford, Alec and Kim, Jong Wook and Xu, Tao and Brockman, Greg and McLeavey, Christine and Sutskever, Ilya},
  booktitle={International Conference on Machine Learning (ICML)},
  pages={28492--28518},
  year={2023}
}

@inproceedings{xu2025publicspeak,
  title={{PublicSpeak}: Hearing the Public with a Probabilistic Framework},
  author={Xu, Tianliang and Brown, Eva Maxfield and Dwyer, Dustin and Tomkins, Sabina},
  booktitle={Proceedings of the Thirty-Ninth AAAI Conference on Artificial Intelligence (AAAI-25)},
  pages={28520--28529},
  year={2025}
}

@article{yin2026speakerlm,
  title={{SpeakerLM}: End-to-End Versatile Speaker Diarization and Recognition with Multimodal Large Language Models},
  author={Yin, Han and Chen, Yafeng and Deng, Chong and Cheng, Luyao and Wang, Hui and Tan, Chao-Hong and Chen, Qian and Wang, Wen and Li, Xiangang},
  journal={arXiv preprint arXiv:2508.06372},
  year={2025}
}

@inproceedings{sun2024itercot,
  title={Enhancing Chain-of-Thoughts Prompting with Iterative Bootstrapping in Large Language Models},
  author={Sun, Jiashuo and Luo, Yi and Gong, Yeyun and Lin, Chen and Shen, Yelong and Guo, Jian and Duan, Nan},
  booktitle={Findings of the Association for Computational Linguistics: NAACL 2024},
  pages={4074--4101},
  year={2024}
}

@inproceedings{bredin2023pyannote,
  title={pyannote.audio 2.1 speaker diarization pipeline: principle, benchmark, and recipe},
  author={Bredin, Herv{\'e}},
  booktitle={Proc. Interspeech},
  year={2023}
}

@article{park2022review,
  title={A Review of Speaker Diarization: Recent Advances with Deep Learning},
  author={Park, Tae Jin and Kanda, Naoyuki and Dimitriadis, Dimitrios and Han, Kyu J. and Watanabe, Shinji and Narayanan, Shrikanth},
  journal={Computer Speech \& Language},
  volume={72},
  pages={101317},
  year={2022}
}

@article{gemini2024,
  title={Gemini: A Family of Highly Capable Multimodal Models},
  author={{Gemini Team, Google}},
  journal={arXiv preprint arXiv:2312.11805},
  year={2023}
}

@article{grimmer2013text,
  title={Text as Data: The Promise and Pitfalls of Automatic Content Analysis Methods for Political Texts},
  author={Grimmer, Justin and Stewart, Brandon M.},
  journal={Political Analysis},
  volume={21},
  number={3},
  pages={267--297},
  year={2013}
}

@article{dietrich2019pitch,
  title={Pitch Perfect: Vocal Pitch and the Emotional Intensity of Congressional Speech},
  author={Dietrich, Bryce J. and Hayes, Matthew and O'Brien, Diana Z.},
  journal={American Political Science Review},
  volume={113},
  number={4},
  pages={941--962},
  year={2019}
}

@article{rask2025partisan,
  title={Partisan Conflict in Nonverbal Communication},
  author={Rask, Mathias and Hjorth, Frederik},
  journal={Political Science Research and Methods},
  year={2025}
}

@article{dietrich2019emotional,
  title={Emotional Arousal Predicts Voting on the {U.S. Supreme Court}},
  author={Dietrich, Bryce J. and Enos, Ryan D. and Sen, Maya},
  journal={Political Analysis},
  volume={27},
  number={2},
  pages={237--243},
  year={2019}
}

@article{knox2021dynamic,
  title={A Dynamic Model of Speech for the Social Sciences},
  author={Knox, Dean and Lucas, Christopher},
  journal={American Political Science Review},
  volume={115},
  number={2},
  pages={649--666},
  year={2021}
}

@article{mestre2025potential,
  title={Potential and Pitfalls of Audio as Data for Political Research},
  author={Mestre, Rafael and Ryan, Matt},
  journal={Political Analysis},
  year={2025}
}

@article{tarr2023automated,
  title={Automated Coding of Political Campaign Advertisement Videos: An Empirical Validation Study},
  author={Tarr, Alexander and Hwang, JungHwan and Imai, Kosuke},
  journal={Political Analysis},
  volume={31},
  number={4},
  pages={554--574},
  year={2023}
}

@article{martin2026participation,
  title={Participation and Representation in Local Government Speech},
  author={Martin, Olivia and Venugopal, Amar},
  journal={arXiv preprint arXiv:2604.21202},
  year={2026}
}

@article{landis1977measurement,
  title={The Measurement of Observer Agreement for Categorical Data},
  author={Landis, J. Richard and Koch, Gary G.},
  journal={Biometrics},
  volume={33},
  number={1},
  pages={159--174},
  year={1977}
}

@inproceedings{egami2024dsl,
  title={Using Imperfect Surrogates for Downstream Inference: Design-based Supervised Learning for Social Science Applications of Large Language Models},
  author={Egami, Naoki and Hinck, Musashi and Stewart, Brandon M. and Wei, Hanying},
  booktitle={Advances in Neural Information Processing Systems},
  volume={36},
  year={2023}
}

@book{condorcet1785,
  title={Essai sur l'application de l'analyse \`{a} la probabilit\'{e} des d\'{e}cisions rendues \`{a} la pluralit\'{e} des voix},
  author={de Condorcet, Marquis},
  year={1785},
  publisher={Imprimerie Royale},
  address={Paris}
}

@inproceedings{desplanques2020ecapa,
  title={{ECAPA-TDNN}: Emphasized Channel Attention, Propagation and Aggregation in {TDNN} Based Speaker Verification},
  author={Desplanques, Brecht and Thienpondt, Jenthe and Demuynck, Kris},
  booktitle={Interspeech},
  pages={3830--3834},
  year={2020}
}

@inproceedings{baevski2020wav2vec,
  title={wav2vec 2.0: A Framework for Self-Supervised Learning of Speech Representations},
  author={Baevski, Alexei and Zhou, Henry and Mohamed, Abdelrahman and Auli, Michael},
  booktitle={Advances in Neural Information Processing Systems},
  volume={33},
  pages={12449--12460},
  year={2020}
}

@book{pateman1970participation,
  title={Participation and Democratic Theory},
  author={Pateman, Carole},
  year={1970},
  publisher={Cambridge University Press},
  address={Cambridge}
}

@book{mansbridge1983beyond,
  title={Beyond Adversary Democracy},
  author={Mansbridge, Jane J.},
  year={1983},
  publisher={University of Chicago Press},
  address={Chicago}
}

@book{fung2004empowered,
  title={Empowered Participation: Reinventing Urban Democracy},
  author={Fung, Archon},
  year={2004},
  publisher={Princeton University Press},
  address={Princeton, NJ}
}

\appendix

\section{Algorithms}
\begin{algorithm*}[t]
\caption{Temporal Exemplar Selection}\label{alg:temporal}
\begin{algorithmic}[1]
\Require Query speaker $q$ with first-appearance time $t_q$ in meeting $m$; set of speakers $\mathcal{S}_m$ in $m$ (excluding $q$); number of exemplars $k$
\Ensure Exemplar set $\mathcal{E} \subset \mathcal{S}_m$, $|\mathcal{E}| = k$
\For{each speaker $j \in \mathcal{S}_m$}
    \State $t_j \gets$ first-appearance time of $j$ in $m$
    \State $d_j \gets |t_j - t_q|$
\EndFor
\State Sort $\mathcal{S}_m$ by $d_j$ in ascending order
\State $\mathcal{E} \gets$ first $k$ speakers from sorted list
\State \Return $\mathcal{E}$
\end{algorithmic}
\end{algorithm*}

\begin{algorithm*}[t]
\caption{Iterative Bootstrapping}\label{alg:bootstrap}
\begin{algorithmic}[1]
\Require Segments $\mathcal{S} = \{s_1, \ldots, s_N\}$ from meeting $m$; baseline classifier $\mathcal{C}_0$ (5-shot, random exemplars); number of temporal neighbors $k$; maximum passes $T$
\Ensure Final label assignment $\mathbf{L}_T$
\State $\mathbf{L}_0 \gets \mathcal{C}_0(\mathcal{S})$ \Comment{Pass 0: baseline classification}
\For{$t = 1$ to $T$}
    \For{each segment $s_i \in \mathcal{S}$}
        \State $\mathcal{E}_i \gets \textsc{TemporalExemplars}(s_i,\, k)$ \Comment{Alg.~\ref{alg:temporal}}
        \For{each exemplar $e \in \mathcal{E}_i$}
            \State Assign label $\mathbf{L}_{t-1}(e)$ to exemplar $e$
        \EndFor
        \State $\mathbf{L}_t(s_i) \gets \textsc{Classify}(s_i,\, \mathcal{E}_i)$
    \EndFor
    \If{$\mathbf{L}_t = \mathbf{L}_{t-1}$} \textbf{break} \Comment{Convergence}
    \EndIf
\EndFor
\State \Return $\mathbf{L}_T$
\end{algorithmic}
\end{algorithm*}

\section{Additional Validation Detail}
\label{sec:appendix_validation}

\subsection{Bootstrapping Convergence}
Table~\ref{tab:convergence} reports per-pass performance of the bootstrapping procedure on the production ground truth ($N = 1{,}749$--$1{,}761$ speakers across five cities); Pass~0 is a zero-shot classification and subsequent passes use the previous pass's predicted labels for temporal exemplar selection. Temporal context systematically shifts the error pattern: the zero-shot baseline has lower public recall (0.722), which bootstrapping raises to 0.792 by Pass~3 while holding public precision near-constant (0.869 to 0.859). Across the three passes the procedure is net-corrective---it fixes 190 previously-incorrect labels while newly breaking 157 (net $+33$), and the corrections disproportionately rescue true members of the public (127 of 190). The regressions are roughly balanced across classes (80 true officials, 77 true members of the public), and we detect a modest within-meeting propagation effect: speakers whose nearest labeled neighbors carried an incorrect prior-pass label regress about twice as often as those who do not (3.7\% vs.\ 1.6\%), which---together with a near-neutral final pass---is why we stop at $T=3$. Absolute label churn is far larger over the full corpus (7,826 cumulative changes across the three passes over all 29,251 production speakers), reflecting the unlabeled population at scale rather than the labeled evaluation set.

\begin{table*}[t]
\centering
\caption{Iterative Bootstrapping Convergence on Production Ground Truth ($k = 4$ temporal neighbors, Gemini 2.5 Flash, $N \approx 1{,}761$, 5 cities).}
\label{tab:convergence}
\begin{tabular}{lccccccc}
\toprule
Pass & $N$ & Acc. & $\kappa$ & Pub.\ P & Pub.\ R & Pub.\ F1 \\
\midrule
0\ \ (zero-shot)               & 1,759 & 0.848 & 0.672 & 0.869 & 0.722 & 0.789 \\
1\ \ (temporal + Pass 0 labels)& 1,749 & 0.864 & 0.711 & 0.862 & 0.781 & 0.819 \\
2\ \ (temporal + Pass 1 labels)& 1,759 & \textbf{0.869} & \textbf{0.721} & 0.864 & 0.790 & \textbf{0.826} \\
3\ \ (temporal + Pass 2 labels)& 1,761 & 0.867 & 0.718 & 0.859 & \textbf{0.792} & 0.824 \\
\bottomrule
\end{tabular}
\end{table*}

\begin{figure*}[t]
\centering
\resizebox{\linewidth}{!}{%
\begin{tikzpicture}[
    node distance=0.7cm and 1.0cm,
    box/.style={rectangle, draw, rounded corners=3pt, minimum width=2.0cm,
                minimum height=0.85cm, align=center, font=\small},
    pass/.style={box, fill=blue!8},
    lbl/.style={box, fill=orange!10, minimum width=1.4cm},
    proc/.style={box, fill=green!8, minimum width=2.2cm},
    arrow/.style={-{Stealth[length=2.5mm]}, thick},
    ann/.style={font=\scriptsize}
]

\node[pass] (p0) {Pass 0\\(5-shot)};
\node[lbl, right=of p0] (l0) {$\mathbf{L}_0$};
\draw[arrow] (p0) -- (l0);

\node[proc, right=of l0] (ts1) {Temporal\\selection};
\draw[arrow] (l0) -- (ts1);

\node[pass, right=of ts1] (p1) {Pass 1};
\node[lbl, right=of p1] (l1) {$\mathbf{L}_1$};
\draw[arrow] (ts1) -- (p1);
\draw[arrow] (p1) -- (l1);

\node[right=0.2cm of l1, font=\large] (dots) {$\cdots$};

\node[pass, right=0.2cm of dots] (pn) {Pass $T$};
\node[lbl, right=of pn] (ln) {$\mathbf{L}_T$};
\draw[arrow] (dots) -- (pn);
\draw[arrow] (pn) -- (ln);

\node[ann, below=0.2cm of ln] {converged};

\draw[arrow, densely dashed, blue!70]
    (l1.south) -- ++(0,-0.6) -| (ts1.south);

\end{tikzpicture}
}
\caption{Bootstrapping loop.  Pass~0 classifies segments with the baseline 5-shot
classifier.  Subsequent passes use the previous pass's predicted labels as
exemplar labels for temporal selection, iterating until convergence.}
\label{fig:bootstrap}
\end{figure*}
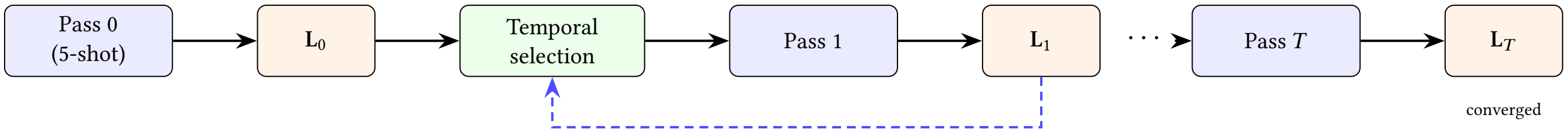

\begin{figure*}[t]
\centering
\resizebox{\linewidth}{!}{%
\begin{tikzpicture}[
    seg/.style={rectangle, draw, minimum width=1.2cm, minimum height=0.7cm,
                align=center, font=\small},
    win/.style={rectangle, draw, densely dashed, rounded corners=2pt,
                inner sep=4pt},
    arrow/.style={-{Stealth[length=2mm]}, thick},
    ann/.style={font=\scriptsize, align=center}
]

\foreach \i/\lab in {1/$s_1$, 2/$s_2$, 3/$s_3$, 4/$s_4$, 5/$s_5$} {
    \pgfmathsetmacro{\x}{(\i-1)*1.6}
    \node[seg] (s\i) at (\x, 0) {\lab};
}

\node[win, fill=blue!5, fit=(s1)(s3), label={[ann]above:Agent $a_2$}] (wA) {};
\draw[arrow, blue!70] (wA.south) -- ++(0,-0.5) node[below, ann] {classify $s_2$};

\foreach \i/\lab in {1/$s_1$, 2/$s_2$, 3/$s_3$, 4/$s_4$, 5/$s_5$} {
    \pgfmathsetmacro{\x}{(\i-1)*1.6}
    \node[seg] (t\i) at (\x, -2.2) {\lab};
}

\node[win, fill=red!5, fit=(t2)(t4), label={[ann]above:Agent $a_3$}] (wB) {};
\draw[arrow, red!70] (wB.south) -- ++(0,-0.5) node[below, ann] {classify $s_3$};

\foreach \i/\lab in {1/$s_1$, 2/$s_2$, 3/$s_3$, 4/$s_4$, 5/$s_5$} {
    \pgfmathsetmacro{\x}{(\i-1)*1.6}
    \node[seg] (u\i) at (\x, -4.4) {\lab};
}

\node[win, fill=green!5, fit=(u3)(u5), label={[ann]above:Agent $a_4$}] (wC) {};
\draw[arrow, green!60!black] (wC.south) -- ++(0,-0.5) node[below, ann] {classify $s_4$};

\end{tikzpicture}
}
\caption{Sliding-window classification.  Each agent receives three consecutive
speaker segments and classifies the middle speaker.  Adjacent windows overlap
by two segments, enabling detection of local turn-taking patterns.}
\label{fig:window}
\end{figure*}
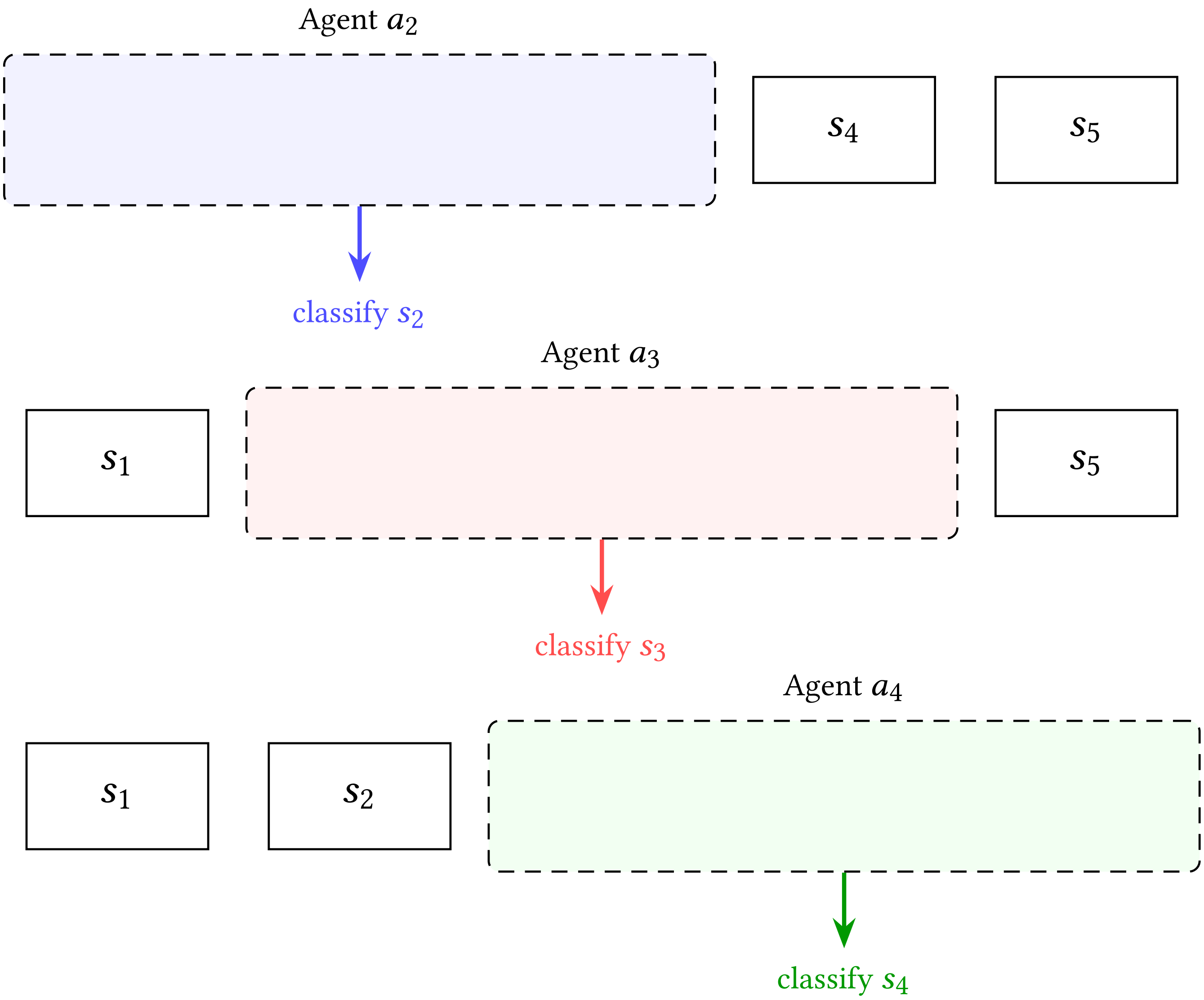

\begin{table*}[t]
\centering
\caption{Supervised Baselines vs.\ MLLM Methods (grouped 5-fold CV on $N = 1{,}081$ ground-truth speakers, 5 cities)}
\label{tab:prosodic}
\begin{tabular}{llccccc}
\toprule
 & Method & Acc. & $\kappa$ & Cit.\ P & Cit.\ R & Cit.\ F1 \\
\midrule
\multicolumn{7}{l}{\textit{Prosodic features (79 features from diarization output)}} \\
& Logistic Regression       & 0.714 & 0.410 & 0.629 & 0.678 & 0.652 \\
& Gradient Boosting         & 0.803 & 0.580 & 0.784 & 0.694 & 0.736 \\
\midrule
\multicolumn{7}{l}{\textit{Speaker embeddings}} \\
& ECAPA-TDNN + Random Forest  & 0.774 & 0.510 & 0.772 & 0.610 & 0.681 \\
& Wav2Vec2 + Logistic Regression & 0.750 & 0.469 & 0.704 & 0.638 & 0.669 \\
\midrule
\multicolumn{7}{l}{\textit{MLLM (Gemini 2.5 Flash, no training data)}} \\
& 5-shot (fixed exemplars)  & 0.853 & 0.683 & 0.881 & 0.727 & 0.796 \\
& Majority vote (3 methods) & \textbf{0.893} & \textbf{0.772} & \textbf{0.906} & \textbf{0.813} & \textbf{0.857} \\
\bottomrule
\end{tabular}
\begin{flushleft}
\footnotesize\textit{Note:} All supervised baselines use grouped 5-fold cross-validation stratified by label, with meetings as groups to prevent leakage.  ECAPA-TDNN embeddings are 192-dimensional speaker verification vectors from a model pretrained on VoxCeleb; Wav2Vec2 embeddings are 768-dimensional mean-pooled hidden states from a model pretrained on LibriSpeech.  Neither embedding model was fine-tuned on our data.
\end{flushleft}
\end{table*}

\paragraph{Statistical comparison.}  We assess pairwise performance differences using McNemar's test with continuity correction. Pass~3 significantly outperforms 5-shot ($\chi^{2} = 5.87$, $p = 0.015$), confirming the value of temporal bootstrapping over fixed exemplars. The majority-vote ensemble significantly outperforms both 5-shot ($\chi^{2} = 25.01$, $p < 0.001$) and window~2b ($\chi^{2} = 15.61$, $p < 0.001$), but does not significantly improve over Pass~3 alone ($\chi^{2} = 2.75$, $p = 0.097$).  The Pass~3 vs.\ window~2b comparison is likewise non-significant ($p = 0.082$), consistent with their similar overall accuracy.

\paragraph{Error complementarity.}  On the 1,761 ground-truth speakers, Pass~3 makes 234 errors, 5-shot 269, and window~2b 260.  Critically, a substantial share of each method's errors are unique: 24\% of Pass~3 errors are not made by either other method, compared to 29\% for 5-shot and 30\% for window~2b.  Pairwise inter-method agreement (Cohen's $\kappa$) ranges from 0.725 (5-shot vs.\ window~2b) to 0.751 (5-shot vs.\ Pass~3), indicating that the methods agree substantially but leave room for ensemble gains.  The high proportion of unique errors---despite all three methods using the same underlying MLLM---reflects their distinct input representations (fixed exemplars, temporal neighbors, sliding window) and explains why majority voting improves over any individual method.

\subsection{Robustness Across Speaker Characteristics}
\label{sec:robustness}

A potential concern is that classification accuracy may vary systematically with speaker characteristics, introducing bias into downstream measurements. We examine robustness along two dimensions: total speaking duration and temporal entry position.

\paragraph{Speaking duration.}  We bin ground-truth speakers by total speaking time from the diarization output (not the clips sent to the MLLM). The majority-vote ensemble performs consistently across duration bins: 87.5\% for speakers with $<$30\,s of total speaking time ($N = 312$), 89.2\% for 30--120\,s ($N = 574$), 84.9\% for 120--300\,s ($N = 511$), and 89.3\% for $>$300\,s ($N = 364$). The lack of a monotone relationship between duration and accuracy suggests that the pipeline does not systematically favor verbose speakers over brief ones---important because members of the public tend to speak less than officials.

\paragraph{Temporal position.}  We compute each speaker's relative entry position (first appearance divided by meeting length) and bin speakers into tertiles. The majority-vote ensemble achieves 91.8\% accuracy for early speakers (first third of the meeting), 86.0\% for middle, and 85.2\% for late. The higher accuracy for early speakers reflects the concentration of officials at the start of meetings during procedural business, where the classification signal is strongest. Late-meeting speakers---predominantly members of the public during public comment---are classified less accurately but still at 85.2\%, well above the 60.7\% majority-class baseline. All three individual methods show the same monotone pattern, confirming that temporal position affects difficulty rather than introducing method-specific artifacts.

We compare against two families of supervised baselines. Because these baselines require extracting acoustic features and speaker embeddings from source audio, we evaluate them---and, for a like-for-like comparison, the MLLM ensemble---on the $N = 1{,}081$-speaker subset (annotation rounds R2, R4--R7) for which those features were computed.  The first family uses \emph{hand-crafted prosodic features}: numeric summaries of the sound itself---how high or low the voice is (pitch), how loud (energy), how the sound's frequency content is distributed (spectral measures and MFCCs, standard descriptors of voice timbre)---together with where and how much the speaker talks in the meeting (first appearance, entry position, total duration, segment count), 79 features in all.  The second family uses \emph{pretrained neural speaker embeddings}: numeric voice ``fingerprints'' produced by deep models built to recognize \emph{individual} voices---ECAPA-TDNN \citep{desplanques2020ecapa}, a 192-dimensional speaker-verification model trained on VoxCeleb, and Wav2Vec2 \citep{baevski2020wav2vec}, a 768-dimensional self-supervised speech model trained on LibriSpeech.  For each speaker, we extract embeddings from up to three clips (first, middle, last by start time) and average across clips.  All supervised methods use grouped 5-fold cross-validation with meetings as groups to prevent leakage. We move the full feature list and importance rankings to Appendix~\ref{sec:appendix_tables}.

The speaker embedding baselines are informative about what audio representations capture.  ECAPA-TDNN, a model designed for speaker \emph{verification} (distinguishing individual identities), achieves only 77.4\%---better than chance but far below the MLLM.  This indicates that speaker role is not simply a function of voice identity: officials and members of the public span diverse vocal characteristics.  Wav2Vec2, a general-purpose speech representation model, performs similarly (75.0\%), suggesting that self-supervised speech features also do not capture the higher-order cues that distinguish roles.

Feature importance analysis of the prosodic classifiers reveals that temporal position (\texttt{first\_appearance}, \texttt{rel\_entry}) dominates acoustic features, consistent with our finding that members of the public systematically enter meetings later than officials (Section~\ref{sec:results}; full rankings in Appendix~\ref{sec:appendix_tables}).  This suggests that prosodic features capture low-level acoustic properties that overlap substantially between officials and members of the public---both are adults speaking into microphones in the same room---while the MLLM can additionally leverage higher-order cues: procedural vocabulary spoken with practiced fluency, turn-taking rhythm with the chair, and contextual signals that are audible but not captured by summary statistics of the acoustic signal.

\paragraph{Cross-city generalization.}  A leave-one-city-out (LOCO) evaluation on the same $N = 1{,}081$-speaker feature-extraction subset reveals a sharp contrast in generalizability. For the prosodic Gradient Boosting classifier, training on four cities and testing on the held-out city yields a macro-average accuracy of 70.0\%, compared to 80.3\% under within-city cross-validation---a 10-point drop. Performance varies widely by city (from 61.2\% in New Brunswick to 80.6\% in Jersey City), reflecting the extent to which temporal and acoustic features are city-specific. The MLLM majority-vote ensemble, by contrast, achieves a macro-average accuracy of 89.5\% across the same five LOCO folds, with a narrow range from 87.6\% (Asbury Park) to 91.8\% (Brockton). The MLLM's ability to generalize without retraining is a key advantage for scaling to new cities: it requires no labeled data from the target jurisdiction, whereas a supervised prosodic classifier trained on other cities would suffer substantial degradation. We caution, however, that this is generalization across \emph{similar} jurisdictions: all five cities are urban or suburban, in Democratic-leaning states, and hold meetings in a broadly comparable format. The held-out city in each LOCO fold differs in council size and acoustics but not in basic institutional type. Whether the pipeline transfers to systematically different contexts---rural town meetings, partisan-charged hearings, or jurisdictions with very different public-comment rules---is not established here and is a priority for future validation (Section~\ref{sec:discussion}).



\section{Data Collection Details}
\label{sec:appendix_scraping}

\subsection{Audio Acquisition}

Meeting recordings were obtained from official city YouTube channels using \texttt{yt-dlp}, an open-source command-line tool for downloading media from YouTube and other platforms.  For each of the seven sample cities, we identified the official government YouTube channel and downloaded all available city council meeting recordings posted between January~2017 and December~2023.  The download command extracted the best available audio stream directly (without re-encoding video), producing MP3 files at the source bitrate (typically 128--192\,kbps).  Meeting titles, dates, and URLs were logged in a metadata CSV for provenance tracking.

\subsection{Audio Format Conversion}

Raw MP3 files were converted to 16\,kHz mono WAV format using \texttt{librosa} prior to diarization, as the \texttt{pyannote.audio} pipeline expects single-channel input at 16\,kHz.  Original MP3 files were retained on Dropbox as the canonical source; WAV conversion was performed on-the-fly during batch processing to minimize local storage requirements.  Individual recordings range from approximately 30~minutes to over 5~hours in duration.

\subsection{Storage and Reproducibility}

The data pipeline is organized into four tiers:
\begin{enumerate}[nosep]
    \item \textbf{Raw audio} (MP3 files, ${\sim}$2\,TB total) is stored on Dropbox and is not included in the code repository due to file size.
    \item \textbf{Diarization output} (segment-level CSVs with timestamps and prosodic features, ${\sim}$940\,MB) is stored locally.
    \item \textbf{Sampled audio clips} (short WAV files, one per speaker segment) are generated on-the-fly for classification and are treated as ephemeral intermediate files.
    \item \textbf{Classification results} (speaker-level labels, small CSVs) are the final research data and are version-controlled in the project repository.
\end{enumerate}
All processing scripts are included in the project repository.  Diarization and classification outputs are linked to source recordings by filename, enabling end-to-end replication given access to the original YouTube URLs.

\subsection{Classification Model and API Settings}
\label{sec:appendix_repro}

\paragraph{Model and interface.} All three production classification strategies (5-shot, Pass~3 bootstrapping, and window~2b) use Google \texttt{gemini-2.5-flash} through the \texttt{google-genai} Python client, calling \texttt{client.models.generate\_content}. Audio clips are transmitted with the Gemini Files API (\texttt{client.files.upload}). Supervised and modality baselines use different backbones (\texttt{gemini-2.5-flash-lite} for some ablations, \texttt{whisper-large-v3} for ASR, \texttt{Qwen2-Audio-7B}, ECAPA-TDNN, and Wav2Vec2), reported where used.

\paragraph{Generation settings and determinism.} Every call sets \texttt{temperature}~$=0.0$; we set no custom \texttt{top\_p}, \texttt{top\_k}, or \texttt{seed}, and use the API's default safety settings (no custom \texttt{HarmCategory} thresholds). Temperature~0 makes each classification deterministic given identical input, so repeated runs reproduce the same labels and the bootstrapping loop converges to a fixed point rather than drifting on run-to-run randomness. Upstream clip selection is seeded (\texttt{seed}~$=42$) with a 30\,s-per-speaker cap in weighted sampling.

\paragraph{Failures, retries, and unparsable responses.} Each API call is retried up to three times (four for the ASR baseline) with exponential back-off; rate-limit errors (HTTP~429 / \texttt{RESOURCE\_EXHAUSTED}) trigger back-off-and-retry rather than dropping the speaker. The text response is lowercased and scanned for ``official'' or ``citizen''; if neither token is present---or the response is empty or blocked by a safety filter---the clip is labeled \texttt{unknown} and excluded from evaluation, and a call that still errors after all retries is recorded as an error row. Under the 5-shot~v3 prompt, $\geq$95\% of responses are parseable.

\paragraph{Run provenance and scale.} The production classification was run in \textbf{May~2026} (bootstrapping passes~0--3 completed 2026-05-10; the 5-shot and window~2b runs completed 2026-05-18). Across the three strategies the run issued \textbf{175{,}493 model calls} (117{,}004 for bootstrapping across passes~0--3; 29{,}238 for 5-shot; 29{,}251 for window~2b), consuming approximately \textbf{309~million input tokens} (audio tokenization dominates this count) and \textbf{168{,}000 output tokens} (classifications are a single word). At list Gemini~2.5~Flash rates this corresponds to an API cost on the order of \textbf{\$90}. \todo{confirm the final billed API cost and the exact per-token audio/text rates in effect during the May 2026 run.}

\paragraph{Artifacts.} Code (diarization, sampling, prompts, classification, and analysis scripts), releasable annotations, split files, and the exemplar pool are available from the authors on reasonable request. Per-call token usage and predicted labels for every pass are stored in the \texttt{output/production\_results*} CSVs, from which the counts above are reproduced.

\subsection{Corpus Documentation (Data Statement)}
\label{sec:appendix_corpus_doc}

Table~\ref{tab:datastatement} documents the corpus as a data statement, consolidating purpose, provenance, scale, preprocessing, partitions, leakage controls, known biases, and access terms in one place.

\begin{table*}[t]
\centering
\caption{Corpus documentation / data statement.}
\label{tab:datastatement}
\footnotesize
\renewcommand{\arraystretch}{1.15}
\begin{tabularx}{\textwidth}{@{}p{0.20\textwidth}X@{}}
\toprule
\textbf{Dimension} & \textbf{Description} \\
\midrule
Purpose & Measure public participation (speaker role and floor-time share) in U.S.\ local-government meetings. \\
Sources & Official city YouTube channels and government websites; automated scraping (Section~\ref{sec:appendix_scraping}). \\
Dates & Meetings recorded 2017--2023. \\
Collection frame vs.\ analytic sample & Frame: 1,618 downloaded recordings from 7 cities (Table~\ref{tab:cities}). Analytic sample: 1,565--1,567 diarized meetings from 5 production cities (Table~\ref{tab:corpus_desc}). \\
Exclusions & Oakland (corrupted / too short for diarization); East Palo Alto (development only); corrupted downloads, files below the minimum-duration threshold, and diarization failures (51-meeting frame-to-analytic reduction). \\
Scale and units & $\approx$2,600 hours; 199,118 speech segments; 29,257 meeting-level speakers. Units: meeting, speaker, segment. \\
Formats and preprocessing & MP3/WAV audio resampled to 16\,kHz mono; segments $<0.5$\,s discarded; weighted sampling caps each speaker at 30\,s (Section~\ref{sec:method}). \\
Annotations & Speaker role (official / public participant) by three research assistants over eight rounds; disagreements resolved by majority vote (Section~\ref{sec:ground_truth_overview}). \\
Data partitions & Development (Albany, Chino; prompt tuning and diarization checks); validation ground truth (5 production cities, $N=1{,}761$); production corpus (unlabeled, application only). See Table~\ref{tab:partitions}. \\
Leakage controls & Development cities are disjoint from production; the production corpus is held out from prompt, exemplar, pass, and ensemble selection. \\
Missingness & Uneven temporal coverage (2019--2022 dominant); video framing present for 45--100\% of speakers by city; ASR / open-audio comparisons run on subsets with locally available source audio. \\
Known biases & Purposive sample of dense, urban/suburban, Democratic-leaning cities drawn from a rent-control frame; YouTube-posting requirement favors higher-capacity clerk offices. Not a representative census of U.S.\ local government. \\
Privacy & Public records of open meetings; analysis assigns de-identified labels (SPEAKER\_00, \ldots) and classifies by institutional role, not individual identity. \\
Access and licensing & Original meeting audio is public record at the cited city channels (referenced by pointer, not redistributed); derived annotations and code available from the authors on reasonable request. \\
Intended / prohibited uses & Intended: aggregate measurement of participation and institutional comparison. Prohibited: identifying, profiling, or re-contacting individual speakers. \\
Versioning and citation & Cite this paper for the instrument and corpus. \\
\bottomrule
\end{tabularx}
\end{table*}

\section{MLLM Prompt Variants}
\label{sec:appendix_experiments}

Table~\ref{tab:prompt_variants} reports classification performance across the five prompt configurations we tested on the 402-segment ground truth (Gemini 2.5 Flash, temperature 0). All valid predictions are included for each variant; the $N_\text{valid}$ column reports how many segments received a parseable response.

\begin{table*}[t]
\centering
\caption{MLLM Prompt Variant Performance (Gemini 2.5 Flash, $N_\text{total} = 402$ segments)}
\label{tab:prompt_variants}
\begin{tabular}{llccccc}
\toprule
Variant & Description & $N_\text{valid}$ & Accuracy & $\kappa$ & Cit.\ P & Cit.\ R \\
\midrule
0-shot    & No exemplars; original prompt             & 358 & 86.6\% & 0.554 & 0.63 & 0.65 \\
0-shot v2 & No exemplars; improved system prompt      & 402 & 88.3\% & 0.561 & 0.68 & 0.59 \\
4-shot    & 4 random exemplars; original prompt       &  89 & 89.9\% & 0.682 & 0.65 & 0.87 \\
4-shot v2 & 4 balanced exemplars; refined prompt      & 400 & 85.5\% & 0.599 & 0.54 & 0.93 \\
5-shot v3 & 4 official + 1 public; minimal prompt     & 385 & \textbf{89.9\%} & \textbf{0.683} & \textbf{0.66} & \textbf{0.85} \\
\bottomrule
\end{tabular}
\end{table*}

\noindent Several patterns emerge. First, prompt engineering substantially affects parse reliability: the original 4-shot prompt produced valid responses for only 89 of 402 segments (22.1\%), while all revised variants achieve $\geq$95\% validity. Second, adding exemplars consistently improves $\kappa$ relative to zero-shot, though the gain is modest. Third, the 5-shot v3 configuration---which provides four official exemplars and one public-participant exemplar from previously validated ground truth with a minimal instruction prompt---achieves the best balance of accuracy (89.9\%), reliability ($\kappa = 0.683$), and public recall (0.85), and is used as the baseline for all subsequent experiments.

\subsection{Prompt Text}

Below we reproduce the exact text prompts used in each configuration. In all few-shot variants, labeled audio exemplars are prepended before the query clip; the text prompt follows the final audio input.

\medskip
\noindent\textbf{Minimal prompt} (used in 0-shot, 0-shot v2, 5-shot v3, and all production runs):

\begin{quote}\small\ttfamily
Listen to this audio clip from a city council meeting.\\[4pt]
Classify the speaker as either:\\
- "official" (government official or staff member)\\
- "citizen" (member of the public)\\[4pt]
Respond with ONLY ONE WORD: either "official" or "citizen"\\[4pt]
Response (one word only):
\end{quote}

\noindent The 0-shot and 0-shot v2 variants differ only in the temperature parameter (default vs.\ 0.0), not in prompt wording.

\medskip
\noindent\textbf{Detailed prompt} (used in 4-shot):

\begin{quote}\small\ttfamily
You are an expert at identifying speaker roles in city council meetings.\\[4pt]
LISTEN FOR THESE CLUES:\\[4pt]
OFFICIALS typically:\\
- Use procedural language: "motion to approve", "second the motion", "agenda item X"\\
- Refer to staff reports, legal analysis, budget items\\
- Speak with formal, professional tone\\
- May call on speakers or manage the meeting flow\\
- Direct the meeting proceedings\\
- Say things like "I move to...", "the chair recognizes...", "the clerk will call roll"\\[4pt]
CITIZENS typically:\\
- Address the council: "Mr.\ Mayor", "Council members", "Madam Chair"\\
- Introduce themselves: "My name is...", "I'm a resident of..."\\
- Share personal experiences or concerns\\
- Speak more emotionally or passionately about issues\\
- Ask the council to take action on something\\
- Say things like "I live at...", "My neighborhood...", "I'm here to speak about..."\\
- Sound like they are ADDRESSING the council, not PART OF the council\\[4pt]
CRITICAL DISTINCTION:\\
- Is this person ADDRESSING the council (speaking TO them) → CITIZEN\\
- Is this person PART OF the council (managing the meeting) → OFFICIAL\\[4pt]
After listening, respond with ONLY ONE WORD:\\
- "official" if government official/staff\\
- "citizen" if member of public\\[4pt]
Response (one word only):
\end{quote}

\medskip
\noindent\textbf{Balanced prompt} (used in 4-shot v2):

\begin{quote}\small\ttfamily
You are an expert at identifying speaker roles in city council meetings.\\[4pt]
LISTEN FOR THESE CLUES:\\[4pt]
OFFICIALS typically:\\
- Use procedural language: "motion to approve", "second the motion", "agenda item"\\
- Refer to staff reports, legal analysis, budget items\\
- Manage the meeting flow: call on speakers, call for votes\\
- Speak with formal, professional tone about city business\\
- Say things like "I move to...", "the chair recognizes...", "the clerk will call roll"\\[4pt]
CITIZENS typically:\\
- Address the council: "Mr.\ Mayor", "Council members"\\
- Introduce themselves: "My name is...", "I'm a resident of..."\\
- Share personal experiences or concerns about issues\\
- Ask the council to take action on something\\
- Say things like "I live at...", "My neighborhood...", "I'm here to speak about..."\\[4pt]
Based on what you hear, classify the speaker.\\[4pt]
Respond with ONLY ONE WORD:\\
- "official" if government official/staff\\
- "citizen" if member of public\\[4pt]
Response (one word only):
\end{quote}

\noindent The key design evolution across variants was simplification: the detailed prompt's ``CRITICAL DISTINCTION'' heuristic and the balanced prompt's enumerated cues both underperformed the minimal prompt when paired with audio exemplars, suggesting that the model extracts classification-relevant information more effectively from reference audio than from textual descriptions of acoustic cues.

\section{Diarization Validation Protocol}
\label{sec:appendix_diarization}

\subsection{Evaluation Design}

We evaluated diarization quality using two complementary tasks administered to three research assistants (RAs) who worked independently.

\textbf{Task~1: Speaker Pair Verification} (126 development pairs from three Chino meetings, plus 360 pairs sampled across four production cities).  Each item consists of two audio clips drawn from the same meeting.  Some pairs come from the same model-assigned speaker label (``same-speaker'' pairs) and some from different labels (``different-speaker'' pairs).  RAs judged whether each pair sounds like the same person, without knowing the model's assignment.  This design detects two error types:
\begin{itemize}[nosep]
    \item \emph{Merge errors:} The model grouped segments from different people under one speaker label.  Detected when RAs judge a same-speaker pair as ``different person.''
    \item \emph{Split errors:} The model assigned one person two different labels.  Detected when RAs judge a different-speaker pair as ``same person.''
\end{itemize}

\textbf{Task~2: Clip Quality Check} (45~clips).  RAs assessed individual clips for single-speaker presence (does the clip contain exactly one voice?), boundary quality (are segment start and end points clean?), and speech audibility (is there intelligible speech?).

\subsection{Answer Key Construction}

The answer key for Task~1 is derived from the diarization model's own assignments: a ``same-speaker'' pair consists of two segments assigned the same speaker label by \texttt{pyannote}, and a ``different-speaker'' pair consists of segments from two different labels.  The key assumption is that the model's assignments are the ground truth to be validated.  When RAs disagree with the model, this flags a potential model error.  Pairs were sampled to include roughly equal numbers of same-speaker and different-speaker items, drawn from speakers across the full duration of each meeting.

\subsection{Inter-Rater Reliability}

Table~\ref{tab:irr_diarization} reports pairwise inter-rater agreement for the Chino development meetings, where average Cohen's $\kappa = 0.897$ (``almost perfect''). Across the four production cities agreement is lower but still substantial (mean pairwise $\kappa = 0.79$, range $0.69$--$0.84$), consistent with their noisier audio; pooling all five cities gives a mean pairwise $\kappa = 0.84$. In each case the reliability confirms that the verification task is well-defined and that RAs can distinguish speakers by voice.

\begin{table}[t]
\centering
\caption{Inter-Rater Reliability for Diarization Verification, Chino Development Meetings (Pairwise Cohen's $\kappa$)}
\label{tab:irr_diarization}
\begin{tabular}{lcc}
\toprule
RA Pair & Cohen's $\kappa$ & Raw Agreement \\
\midrule
RA\,1 vs.\ RA\,2 & 1.000 & 100.0\% \\
RA\,1 vs.\ RA\,3 & 0.902 & 95.2\% \\
RA\,2 vs.\ RA\,3 & 0.787 & 89.5\% \\
\midrule
\textbf{Average} & \textbf{0.897} & \\
\bottomrule
\end{tabular}
\end{table}

\subsection{Error Analysis}

In the Chino development set, all five pairs flagged unanimously by all three RAs as model errors are merge errors with a consistent root cause: the audio clip contains overlapping speech from multiple people, but the model assigned the entire segment to a single speaker label.  These are segmentation boundary errors rather than speaker embedding failures---the model correctly identifies speakers' voices but sometimes draws segment boundaries too broadly, capturing speaker transitions within a single segment.  No split errors were detected in Chino by majority vote (0 of 45 different-speaker pairs flagged as same person).  The four production cities confirm this pattern: merge errors remain the dominant failure mode (12.2\% of same-speaker pairs by majority vote), while split errors---absent in Chino---appear only at a low rate (7.3\% of different-speaker pairs) and are likewise concentrated in overlapping or degraded audio.

\subsection{Clip Quality}

Majority-vote quality assessments: 86.7\% of clips contain a single speaker, and 97.8\% contain audible speech.  Boundary quality showed the most inter-rater variation (66.7--73.3\% rated as ``clean''), consistent with exact cut-off points being a more subjective judgment.

\subsection{Diarization Error Rate (DER and JER)}
\label{sec:appendix_der}

Beyond the pairwise verification above, three research assistants densely annotated speaker turns for seven complete meetings across the production cities (five double-annotated), correcting the pyannote output in Audacity; we score against pyannote with \texttt{pyannote.metrics}. Table~\ref{tab:der} reports per-meeting DER and JER. Errors concentrate almost entirely in Brockton (DER 17--22\%), whose corpus is dominated by school-committee and subcommittee sessions with noisier, overlapping audio; the council-style meetings in the other cities fall at or below 6.8\% DER (3.1\% excluding Brockton). Brockton nonetheless attains among the \emph{highest} role-classification accuracy (90.8\%): because the classifier aggregates a majority vote over several sampled clips per speaker, diarization errors---including the speaker splits and merges that inflate JER---are absorbed at the clip level rather than propagated to the speaker-role labels. Two caveats bound these values: annotators corrected the pyannote draft rather than labeling from scratch, so DER is a mild lower bound, and inter-annotator DER is 15.2\%, reflecting the difficulty of exact boundary placement on overlapping speech.

\begin{table}[t]
\centering
\caption{Diarization Error Rate (DER) and Jaccard Error Rate (JER) on seven densely-annotated meetings (0.25\,s collar, annotator-averaged). Errors concentrate in Brockton's subcommittee audio.}
\label{tab:der}
\begin{tabular}{lcc}
\toprule
Meeting & DER & JER \\
\midrule
Jersey City (Special, Sep.\ 2021)  & 0.4\%  & 1.9\%  \\
Asbury Park (Dec.\ 2018)           & 1.3\%  & 3.9\%  \\
Asbury Park (Special, Mar.\ 2023)  & 3.9\%  & 11.1\% \\
Jersey City (Nov.\ 2021)           & 4.6\%  & 28.0\% \\
Inglewood (Feb.\ 2022)             & 6.8\%  & 23.8\% \\
Brockton (Dec.\ 2019)              & 16.8\% & 50.7\% \\
Brockton (May 2020)                & 21.7\% & 39.9\% \\
\midrule
\textbf{Speech-weighted} & \textbf{7.6\%} & \textbf{21.9\%} \\
\bottomrule
\end{tabular}
\end{table}

\section{Ground Truth Creation}
\label{sec:appendix_ground_truth}

\subsection{Annotation Protocol}

\textbf{What the RAs did.} Three research assistants (RAs) independently labeled each speaker as ``official,'' ``citizen,'' or ``other'' (the last reserved for clips with no intelligible speech, pure background noise, or corrupted audio). For each speaker, the RA listened to up to three short clips (2--15~seconds each) drawn from different points in that speaker's turns; a single clip sufficed when the role was obvious, and the additional clips were available when it was not. Alongside each label the RA recorded a confidence rating (high, medium, low) and an optional free-text note (e.g., ``states name and address,'' ``sounds like roll call,'' ``too short to tell''), which the PI used to audit borderline cases.

\textbf{The decision rubric.} RAs were given a written guide instructing them to rely on role-diagnostic cues rather than on the substantive content of speech: procedural and parliamentary language (``I move to approve,'' ``is there a second,'' ``roll call''), forms of address (officials address one another by title; members of the public introduce themselves and address the council), the audio source (dais microphone versus podium or call-in line), and the speaker's position and footprint in the meeting (officials recur throughout and appear early; members of the public typically speak once, briefly, and later). When genuinely uncertain between the two roles, RAs were instructed to apply the public-participant label (recorded as ``citizen'') at low confidence, since members of the public are the rarer and more consequential class to recover.

\textbf{Training and skills required.} The task does not require subject-matter expertise or transcription skill; RAs were onboarded with the written guide and a small set of worked examples. Importantly, because labeling keys on \emph{how} and \emph{when} a person speaks---procedural register, self-introduction, microphone source, meeting timing---rather than on understanding \emph{what} they say, the task is robust to the features that complicate transcription: a speaker's accent, non-native or informal grammar, and partially inaudible content rarely obstruct role identification. The genuine difficulties in the task, reflected in the RAs' notes and low-confidence flags, were short or overlapping clips and the occasional ambiguous staff voice, which the confidence ratings and multi-clip design were intended to absorb. Each RA worked independently without access to the others' labels. Inter-rater reliability was computed using Fleiss' $\kappa$ for all three raters and pairwise Cohen's~$\kappa$; disagreements were resolved via majority vote, with three-way ties adjudicated by the PI through direct listening.

\subsection{Label Propagation}

The consensus label for a speaker is propagated to \emph{all} diarization segments attributed to that speaker within a meeting. Because diarization assigns a consistent speaker ID across a recording, a single human judgment per speaker suffices to label every segment from that speaker. This label propagation step assumes that speaker identity is stable within a meeting---an assumption supported by the low rate of split errors in our diarization validation: none in the Chino development set and roughly 7\% of different-speaker pairs in the production cities (Section~\ref{sec:validation}).

\subsection{Annotation Rounds}

The ground truth was assembled across eleven annotation rounds (R1--R11), with Round~8 excluded from all analyses due to low inter-rater reliability (Fleiss' $\kappa = 0.511$).

\begin{itemize}[nosep]
    \item \textbf{Round~1} (development): 6 speakers from one Albany, NY meeting (February~3, 2020).
    \item \textbf{Round~2} (development + early production): 234 speakers from meetings across Chino (CA) and initial production cities.
    \item \textbf{Rounds~4--7} (production): Extended annotation to five production cities with progressively improved annotation guidelines, yielding 732 additional labeled speakers.
    \item \textbf{Rounds~9--11} (production): Further annotation across the five production cities, adding 818 labeled speakers (R9: 275, R10: 229, R11: 314).
\end{itemize}

\subsection{Inter-Rater Reliability by Round}

Table~\ref{tab:irr_rounds} reports inter-rater reliability for each annotation round. Reliability improved substantially between early rounds (R2, R4: moderate agreement) and later rounds (R5--R7, R9--R11: substantial to almost perfect agreement, apart from R11 at the low end of substantial), reflecting refinement of the annotation guidelines.

\begin{table*}[t]
\centering
\caption{Inter-Rater Reliability by Annotation Round}
\label{tab:irr_rounds}
\begin{tabular}{lccccc}
\toprule
Round & $N$ & Fleiss' $\kappa$ & Interpretation & Mean pairwise $\kappa$ \\
\midrule
R2 & 234 & 0.534 & Moderate     & 0.539 \\
R4 & 166 & 0.525 & Moderate     & 0.521 \\
R5 & 273 & 0.862 & Almost perfect & 0.846 \\
R6 & 199 & 0.744 & Substantial  & 0.745 \\
R7 & 294 & 0.844 & Almost perfect & 0.845 \\
R9 & 275 & 0.868 & Almost perfect & 0.868 \\
R10 & 229 & 0.748 & Substantial  & 0.746 \\
R11 & 314 & 0.662 & Substantial  & 0.664 \\
\midrule
R8 (excl.) & 185 & 0.511 & Moderate & 0.529 \\
\bottomrule
\end{tabular}
\begin{flushleft}
\footnotesize\textit{Note:} R8 is excluded from all validation analyses due to low inter-rater reliability. Mean Fleiss' $\kappa$ across the eight included rounds (R2, R4--R7, R9--R11) is 0.723 (range 0.525--0.868).
\end{flushleft}
\end{table*}

\subsection{Ground Truth Statistics}

After excluding R8 and speakers labeled ``other'' or ``disputed,'' the combined ground truth comprises \textbf{1,824 consensus-labeled speakers} across \textbf{eight annotation sites} (five production cities plus three development sites---Albany, Chino, and East Palo Alto) and \textbf{171 meetings}. Table~\ref{tab:gt_summary} summarizes the distribution by site. The production validation set consists of the $N = 1{,}761$ speakers from the five production cities who received valid predictions from all three classification methods (this is the count that enters Tables~\ref{tab:classification} and~\ref{tab:convergence}); for the production cities the Official and Public columns below report these prediction-valid speakers, so the five-city rows sum to $1{,}761$.

\begin{table*}[t]
\centering
\caption{Ground Truth Summary by City and Label (R1--R11, excluding R8)}
\label{tab:gt_summary}
\begin{tabular}{llrrrr}
\toprule
City & State & Meetings & Official & Public & $N$ (valid) \\
\midrule
Albany          & NY &  1 &  5  &  1  &  --- \\
Chino           & CA &  3 & 38  &  7  &  --- \\
East Palo Alto  & CA &  2 &  7  &  5  &  --- \\
\midrule
Asbury Park     & NJ & 37 & 266 & 184 & 450 \\
Brockton        & MA & 36 & 234 &  49 & 283 \\
Inglewood       & CA & 35 & 193 & 125 & 318 \\
Jersey City     & NJ & 28 & 252 & 282 & 534 \\
New Brunswick   & NJ & 29 & 124 &  52 & 176 \\
\midrule
\textbf{Total}  &    & \textbf{171} & \textbf{1,119} & \textbf{705} & \textbf{1,761} \\
\bottomrule
\end{tabular}
\begin{flushleft}
\footnotesize\textit{Note:} $N$ (valid) counts speakers from production cities with valid predictions from all three classification methods. Albany, Chino, and East Palo Alto are development cities excluded from production evaluation. Official and Public columns report consensus-labeled speakers (speakers labeled ``other'' or ``disputed'' are excluded).
\end{flushleft}
\end{table*}

\noindent The overall public-participant rate in the production validation set is 39.3\% (692/1,761). Later annotation rounds (R5--R7, R9--R11) were deliberately sampled to include more public-comment-heavy sessions, producing a substantially more balanced evaluation set than the initial development data (15.6\% public-participant rate in Chino/Albany).

\section{Propagation of Measurement Uncertainty}
\label{sec:appendix_error_prop}

Our pipeline introduces three sources of measurement error: diarization (89.4\% accuracy), MLLM classification (87.7\% accuracy), and human annotation noise (Fleiss' $\kappa = 0.723$). Using predicted labels directly in downstream analyses can produce biased point estimates and invalid confidence intervals---even at 90\% accuracy, plug-in estimators can exhibit severely undercovered CIs \citep{egami2024dsl}. We therefore formally characterize how classification error propagates to our substantive quantities: public-participant proportions, floor shares, and temporal comparisons.

We ground our analysis in the Design-based Supervised Learning (DSL) framework of \citet{egami2024dsl}, which provides a general theory for using predicted variables in downstream statistical analyses. Our setting maps directly onto this framework: the MLLM predictions are ``surrogate labels'' $Q_i$, the human annotations are ``gold-standard labels'' $Y_i$, and the downstream estimands are aggregate public-participant proportions and cross-group comparisons. The DSL estimator is doubly robust and provides valid confidence intervals without requiring that the classifier be correctly specified. The misclassification correction we apply below is a special case of DSL for prevalence estimation; for regressions and other downstream analyses, the full DSL framework (implemented in the \texttt{dsl} R package) applies.

We correct for classification bias using the confusion matrix from the majority-vote ensemble evaluated on our ground-truth sample ($N = 1{,}761$). Let $\text{TPR} = P(\hat{Y} = \text{public} \mid Y = \text{public}) = 538/692 = 0.777$ and $\text{FPR} = P(\hat{Y} = \text{public} \mid Y = \text{official}) = 63/1{,}069 = 0.059$. The observed (plug-in) public-participant proportion $\hat{\pi}$ relates to the true proportion $\pi$ by:
\begin{equation}
\hat{\pi} = \pi \cdot \text{TPR} + (1 - \pi) \cdot \text{FPR}.
\end{equation}
Inverting yields the bias-corrected estimator:
\begin{equation}
\pi^{*} = \frac{\hat{\pi} - \text{FPR}}{\text{TPR} - \text{FPR}}.
\label{eq:correction}
\end{equation}

Table~\ref{tab:correction} applies this correction to each city. Because $\text{TPR} - \text{FPR} = 0.719 < 1$, the denominator in Equation~\ref{eq:correction} stretches all proportions away from zero, amplifying cross-city differences rather than attenuating them.

\begin{table}[t]
\centering
\caption{Observed vs.\ Bias-Corrected Public-Participant Proportions}
\label{tab:correction}
\begin{tabular}{lccc}
\toprule
City & Observed (\%) & Corrected (\%) & $\Delta$ (pp) \\
\midrule
Asbury Park   & 29.3 & 32.6 & $+3.3$ \\
Brockton      & 13.4 & 10.4 & $-3.0$ \\
Inglewood     & 34.9 & 40.4 & $+5.5$ \\
Jersey City   & 41.4 & 49.4 & $+8.0$ \\
New Brunswick & 28.6 & 31.6 & $+3.0$ \\
\midrule
\textbf{All}  & \textbf{29.9} & \textbf{33.4} & $\boldsymbol{+3.5}$ \\
\bottomrule
\end{tabular}
\end{table}

We construct confidence intervals for $\pi^{*}$ via the delta method. The asymptotic variance of the corrected estimator is:
\begin{equation}
\begin{split}
\text{Var}(\pi^{*}) \approx {}&\frac{1}{(\text{TPR} - \text{FPR})^{2}} \Big[ \frac{\hat{\pi}(1 - \hat{\pi})}{n} \\
&+ (\pi^{*})^{2}\,\frac{\text{TPR}(1 - \text{TPR})}{n_{1}} \\
&+ (1 - \pi^{*})^{2}\,\frac{\text{FPR}(1 - \text{FPR})}{n_{0}} \Big],
\end{split}
\end{equation}
where $n$ is the production sample size, $n_{1} = 692$ and $n_{0} = 1{,}069$ are the ground-truth public-participant and official counts. For the aggregate ($n = 29{,}238$), we obtain $\text{SE} \approx 0.011$ and a 95\% CI of $[31.3\%, 35.5\%]$.

The key substantive comparisons survive correction and are in fact amplified. The COVID-19 participation gap---the difference in public share between the pre- and post-COVID periods---widens from 14.9 percentage points (observed) to 20.7pp (corrected). The cross-city range in public-participant proportions expands from 28.0pp (observed: Brockton 13.4\% to Jersey City 41.4\%) to 39.0pp (corrected: 10.4\% to 49.4\%). Both differences far exceed the CI width, so statistical significance is preserved under the corrected estimator.

Two additional features of our validation design strengthen these conclusions. First, the 87.7\% accuracy is measured end-to-end on diarized speakers: human annotators labeled the same diarized audio segments that the MLLM classified, so diarization errors are already captured in the confusion matrix and require no separate propagation step. Second, human annotation noise ($\kappa = 0.723$) implies that a non-trivial share of ``ground truth'' labels are themselves erroneous. This means reported accuracy is a \textit{lower bound} on true classifier performance---some apparent classification errors are correct predictions penalized by noisy ground truth.

\subsection{Duration-Weighted Correction for Floor Share}
\label{sec:appendix_floorshare_correction}

The correction above adjusts \emph{prevalence} (headcount). Our headline quantity, however, is public \emph{floor share}---the fraction of speaking \emph{time} held by members of the public---which is more sensitive to classification error, since a single misclassified long-speaking official distorts the share far more than a brief speaker. We therefore apply the same inversion (Eq.~\ref{eq:correction}) to the pooled floor share $\hat{s}$, but with error rates estimated on a \emph{speaking-time-weighted} confusion matrix, in which each ground-truth speaker contributes in proportion to its total speaking time. This yields $\text{TPR}_{w} = 0.706$ and $\text{FPR}_{w} = 0.039$ (versus the unweighted $0.777$ and $0.060$): weighting by duration lowers the effective TPR, because longer public turns are somewhat more often mistaken for or merged with official speech. This is direct in the data---public recall by speaking-duration quartile is $72.5\%$, $83.3\%$, $82.6\%$, and $62.6\%$ from shortest to longest, so the long turns that dominate floor time are the hardest to classify correctly, which is exactly the sensitivity the reviewer's concern anticipates and the reason an unweighted correction understates the adjustment. The corrected pooled floor share is $s^{*} = (\hat{s} - \text{FPR}_{w}) / (\text{TPR}_{w} - \text{FPR}_{w})$.

Table~\ref{tab:floorshare_correction} reports observed and corrected pooled floor shares. As with headcount, the correction \emph{amplifies} the public share and the cross-city spread rather than attenuating them: the aggregate rises from 18.7\% to \textbf{22.2\%} (95\% CI $[19.1, 26.3]$, from a meeting-clustered bootstrap that propagates uncertainty in both $\hat{s}$ and the weighted error rates). Crucially, the corrected floor share (22.2\%) remains well below the corrected headcount share (33.4\%; Table~\ref{tab:correction}), so the participation gap---the paper's central finding---survives error correction on \emph{both} sides.

\begin{table}[t]
\centering
\caption{Observed vs.\ duration-weighted bias-corrected \emph{pooled} public floor share.}
\label{tab:floorshare_correction}
\begin{tabular}{lccc}
\toprule
City & Observed (\%) & Corrected (\%) & 95\% CI \\
\midrule
Asbury Park   & 19.2 & 23.0 & $[18.9, 27.8]$ \\
Brockton      &  6.0 &  3.3 & $[0.7, 5.9]$ \\
Inglewood     & 20.6 & 25.1 & $[19.9, 31.4]$ \\
Jersey City   & 25.2 & 31.9 & $[27.0, 38.4]$ \\
New Brunswick & 27.7 & 35.7 & $[29.8, 42.9]$ \\
\midrule
\textbf{All}  & \textbf{18.7} & \textbf{22.2} & $\boldsymbol{[19.0, 26.3]}$ \\
\bottomrule
\end{tabular}
\begin{flushleft}
\footnotesize\textit{Note:} Corrected via $s^{*}=(\hat{s}-\text{FPR}_{w})/(\text{TPR}_{w}-\text{FPR}_{w})$ with duration-weighted $\text{TPR}_{w}=0.706$, $\text{FPR}_{w}=0.039$. 95\% CIs from a meeting-clustered bootstrap propagating uncertainty in both $\hat{s}$ and the weighted error rates. Computed by \texttt{experiments/r4\_floorshare\_correction.py} and \texttt{experiments/more\_robustness.py}.
\end{flushleft}
\end{table}

\subsection{Transportability of the Error Rates}
\label{sec:appendix_transport}

The correction in Equation~\ref{eq:correction} applies a single pair of error rates---$\text{TPR}=0.777$ and $\text{FPR}=0.059$, estimated on the labeled sample---to every city and period in the production corpus. This is valid only under a \emph{transportability} (label-shift) assumption: conditional on a speaker's true role, the probability of misclassification does not depend on the city, the time period, or the audio conditions, so that only the class prevalence $\pi$ varies across strata while $(\text{TPR}, \text{FPR})$ are constant. Under this assumption the stratum-specific corrected prevalence is recovered by applying the same $(\text{TPR},\text{FPR})$ to each stratum's observed $\hat\pi$.

The assumption is substantive and can fail in two ways relevant here. First, \emph{audio conditions shift}: virtual and hybrid meetings after early 2020 change microphone paths and introduce call-in artifacts, which could lower TPR or raise FPR in the pandemic period specifically---biasing the very COVID comparison we most care about. Second, \emph{prevalence-dependent errors}: in extremely official-heavy meetings (e.g., Brockton), the temporal-context components can inherit the majority class, so the effective TPR may be lower where public participation is rarest. Per-city ensemble accuracy is stable (90.8\% Brockton, 89.0\% Inglewood, up to 0.873 public F1 in Jersey City; Section~\ref{sec:validation}), which is consistent with---but does not prove---approximate transportability.

\paragraph{Transportability, tested directly.} We re-estimate the duration-weighted error rates \emph{within} COVID period. They are \emph{not} constant: $\text{TPR}_{w}$ is $0.567$ pre-pandemic (2017--19), $0.772$ during (2020--21), and $0.643$ post (2022--23), so the single-rate assumption is imperfect---consistent with the concern that call-in and virtual audio after early 2020 shift error rates. We therefore recompute the COVID floor-share trajectory using \emph{period-specific} weighted rates. The corrected public floor share falls from $38.0\%$ (95\% CI $[26.7, 52.3]$) pre-pandemic to $16.7\%$ ($[12.7, 22.4]$) during the pandemic, recovering only to $19.8\%$ ($[10.5, 30.6]$) afterward. The pre- and during-pandemic intervals do not overlap, so the COVID decline is robust to correction and, if anything, \emph{larger} than the observed $23.9\%\!\to\!15.7\%$ drop rather than an artifact of pooling error rates across periods. Together with the meeting-clustered intervals reported above and in Table~\ref{tab:classification}, this completes the planned uncertainty analyses---stratified error rates, cluster-robust variance, a duration-weighted floor-share correction, and interval estimates for the headline quantities---all produced by \texttt{experiments/r4\_floorshare\_correction.py} and \texttt{experiments/ci\_bootstrap.py}.

In sum, all three substantive claims in Section~\ref{sec:results} are robust to the pooled misclassification correction, which shifts point estimates by 3--8 percentage points and \textit{amplifies} rather than attenuates cross-group differences; the stratified, clustered, and duration-weighted analyses above will sharpen these intervals and test the transportability assumption directly. The DSL framework \citep{egami2024dsl} provides formal guarantees---consistency, asymptotic normality, and valid confidence intervals---for these and any subsequent regression-based downstream analyses using the predicted labels.


\section{Additional Tables and Figures}
\label{sec:appendix_tables}

\subsection{Prosodic Feature Importance}

Table~\ref{tab:feature_importance} reports the top~10 feature importance scores from the Random Forest classifier trained on the full production corpus (Section~\ref{sec:validation}).  Temporal features---particularly \texttt{first\_appearance} (when the speaker first enters the meeting) and \texttt{rel\_entry} (relative entry position)---dominate acoustic features, consistent with the structural regularity of council meetings: officials speak early during procedural business, while members of the public appear later during public comment periods.  Pure acoustic features (MFCCs, spectral measures) rank considerably lower, reflecting the substantial acoustic overlap between officials and members of the public speaking in the same room under the same recording conditions.

Because Table~\ref{tab:feature_importance} is fit on production-corpus labels produced by the pipeline itself, it describes what this probe relies on, not necessarily what predicts human judgments. Refitting the identical model on the human ground-truth labels ($N = 1{,}762$) leaves the ordering essentially unchanged---\texttt{first\_appearance} ($0.098$) and \texttt{rel\_entry} ($0.072$) remain the top two---so later public entry is a genuine regularity of the labeled data rather than an artifact of model predictions, and reporting it (Section~\ref{sec:results}) is not circular. This predictiveness is nonetheless \emph{bounded}. Table~\ref{tab:temporal_ablation} isolates it: a classifier restricted to the five temporal features reaches only 75.5\% accuracy, and a trivial rule (``official iff the speaker's first appearance is early'') only 67.4\%, against 87.7\% for the MLLM majority-vote ensemble on the same speakers (entry time alone has ROC AUC $0.72$). Because all temporal information is available to these baselines, the ensemble's $12$--$20$ point margin over them is attributable to acoustic and lexical content, not to temporal position.

\begin{table}[t]
\centering
\caption{Random Forest Feature Importance (200 trees, grouped 5-fold CV, $N = 27{,}087$ speakers, 79 features)}
\label{tab:feature_importance}
\begin{tabular}{rlc}
\toprule
Rank & Feature & Importance \\
\midrule
1 & First appearance (temporal) & 0.073 \\
2 & Relative entry position (temporal) & 0.047 \\
3 & Number of segments (temporal) & 0.033 \\
4 & Total speaking duration (temporal) & 0.033 \\
5 & MFCC\textsubscript{5} mean & 0.023 \\
6 & Last appearance (temporal) & 0.015 \\
7 & MFCC\textsubscript{4} mean & 0.015 \\
8 & MFCC\textsubscript{0} mean (std) & 0.014 \\
9 & MFCC\textsubscript{7} mean & 0.014 \\
10 & MFCC\textsubscript{0} std (mean) & 0.013 \\
\bottomrule
\end{tabular}
\begin{flushleft}
\footnotesize\textit{Note:} Features labeled ``(temporal)'' are speaker-level temporal position and activity features; remaining features are segment-level acoustic features aggregated to the speaker level.  The top four features are all temporal, accounting for 18.6\% of total importance.
\end{flushleft}
\end{table}

\begin{table}[t]
\centering
\caption{Temporal position is predictive but bounded. Accuracy on the human ground-truth speakers ($N = 1{,}762$, 165 meetings); Random Forests use grouped 5-fold cross-validation (no meeting leakage) on the same 79-feature representation as Table~\ref{tab:feature_importance}.}
\label{tab:temporal_ablation}
\begin{tabular}{lc}
\toprule
Classifier / rule & Accuracy \\
\midrule
Majority class (predict all ``official'') & 60.7\% \\
Trivial rule: official iff early first appearance & 67.4\% \\
Random Forest, temporal features only (5) & 75.5\% \\
Random Forest, acoustic features only (74) & 74.5\% \\
Random Forest, all features (79) & 79.7\% \\
MLLM majority-vote ensemble (Section~\ref{sec:validation}) & \textbf{87.7\%} \\
\bottomrule
\end{tabular}
\begin{flushleft}
\footnotesize\textit{Note:} The trivial rule applies the single entry-time threshold that maximizes accuracy; entry time alone has ROC AUC $0.72$ for predicting official. Temporal features are first/last appearance, relative entry position, number of segments, and total speaking duration. All temporal information is available to every row, so the ensemble's margin over the temporal-only and trivial baselines reflects acoustic and lexical content rather than temporal position.
\end{flushleft}
\end{table}

\subsection{Corpus Composition by Meeting Type}

Table~\ref{tab:meeting_type} codes each meeting's type from its title. The corpus mixes regular city-council sessions with meetings of other governing bodies, and the mix differs sharply by city. This is most relevant to Section~\ref{sec:results}: only 16.8\% of Brockton's meetings are regular council sessions, which---together with a council format that surfaces little public comment (even Brockton's regular council meetings have a 0.1\% median public floor share)---explains its low aggregate figure.

\begin{table}[t]
\centering
\caption{Corpus composition by meeting type (\% of each city's meetings; coded from meeting titles).}
\label{tab:meeting_type}
\begin{tabular}{lrrrrr}
\toprule
Meeting type & AP & Brk & Ing & JC & NB \\
\midrule
City council (regular) & 42.9 & 16.8 & 95.7 & 49.8 & 32.4 \\
School committee       & ---  & 28.3 & ---  & ---  & 1.1 \\
Committee              & 0.3  & 22.4 & ---  & 0.3  & 1.1 \\
Subcommittee           & ---  & 14.2 & ---  & ---  & --- \\
Board / commission     & 56.9 & 16.4 & ---  & 42.5 & 56.5 \\
Ceremonial / special   & ---  & 0.2  & 3.5  & ---  & --- \\
Other                  & ---  & 1.6  & 0.7  & 7.4  & 8.8 \\
\midrule
Meetings ($N$)         & 364  & 499  & 141  & 299  & 262 \\
\bottomrule
\end{tabular}
\begin{flushleft}
\footnotesize\textit{Note:} AP = Asbury Park, Brk = Brockton, Ing = Inglewood, JC = Jersey City, NB = New Brunswick; East Palo Alto (2 meetings) omitted. Types coded by keyword from meeting titles (\texttt{experiments/w2\_meeting\_type.py}). The median public floor share \emph{within regular council meetings} is 28.7\% (AP), 0.1\% (Brk), 18.4\% (Ing), 19.5\% (JC), and 36.3\% (NB), so Brockton's low overall figure is not merely a composition effect.
\end{flushleft}
\end{table}

\subsection{Classification Accuracy by Meeting}

Table~\ref{tab:by_meeting} reports per-meeting accuracy for the four main classification methods and the best ensemble, illustrating the heterogeneity discussed in Section~\ref{sec:validation}.

\begin{table}[t]
\centering
\caption{Classification Accuracy by Meeting (Gemini 2.5 Flash)}
\label{tab:by_meeting}
\begin{tabular}{lccccc}
\toprule
Meeting & 0-shot & 5-shot & Pass 3 & Window 2b & AND \\
\midrule
Chino Sept 15 & 0.880 & 0.920 & 0.920 & 0.920 & 0.940 \\
Chino Oct 20  & 0.893 & 0.907 & 0.913 & 0.920 & 0.933 \\
Chino Nov 17  & 0.907 & 0.898 & 0.907 & 0.917 & 0.926 \\
Albany Feb 3   & 0.913 & 0.935 & 0.935 & 0.913 & 0.935 \\
\bottomrule
\end{tabular}
\end{table}

\noindent The November~17 Chino meeting is a consistent outlier for bootstrapping methods due to its unusually low public-participant ratio (${\sim}$10\%), which causes temporal neighbors to be overwhelmingly official and biases bootstrapped predictions.  The AND ensemble performs most consistently across meetings.

\section{Audio-as-Data vs.\ Audio-to-Text-to-Data}
\label{sec:appendix_audio_vs_text}

A natural question is whether our audio-direct approach actually buys anything over the conventional pipeline that first transcribes speech and then classifies the text. This appendix tests that directly by holding everything fixed \emph{except the modality}. For each ground-truth speaker we reuse the same diarized clips that the audio pipeline classifies; we then (i)~transcribe each clip with OpenAI's Whisper \citep{radford2023whisper} and (ii)~classify the resulting transcript, either with the same MLLM operating on text (Gemini, with a prompt mirroring the audio prompt) or with a supervised text classifier (TF-IDF features and logistic regression, grouped 5-fold cross-validation by meeting). The audio arm is our production majority-vote ensemble. Because only the input representation differs, any performance gap is attributable to the transcription bottleneck rather than to the classifier or the task.

\paragraph{Scope and caveats.} These results should be read as preliminary, and three caveats bound them. First, \emph{coverage}: this analysis uses the $593$ ground-truth speakers (public-participant rate $0.40$, matching the full sample) whose source recordings were available locally at the time of analysis, not all $1{,}761$. Second, and importantly, \emph{ASR strength}: we transcribe with a mid-sized model (\texttt{whisper-small.en}) rather than the largest available. A stronger ASR model would help the text pipeline most precisely on the hard, noisy clips; these numbers therefore, if anything, \emph{overstate} the audio advantage in magnitude, and the gap below should be read as an upper bound. The qualitative pattern, however, is robust to this choice. Third, we stratify by a cheap acoustic signal-to-noise (SNR) proxy computed directly from each clip, which is correlated with but not identical to transcription error.

\paragraph{Overall.} On the common set of $593$ speakers (Table~\ref{tab:audio_vs_text}), the audio-direct pipeline outperforms both text pipelines: $+3.5$ accuracy points over Whisper$\to$Gemini and $+6.7$ over Whisper$\to$TF-IDF, with markedly higher public \emph{precision} ($0.95$ vs.\ $0.78$). Transcription is therefore not free for this task: routing the signal through text loses information even when the downstream classifier is the same model.

\begin{table*}[t]
\centering
\caption{Audio-as-Data vs.\ Audio-to-Text-to-Data (same $N = 593$ speakers; modality is the only difference; Whisper \texttt{small.en})}
\label{tab:audio_vs_text}
\begin{tabular}{lcccccc}
\toprule
Pipeline & $N$ & Acc.\ & $\kappa$ & Cit.\ P & Cit.\ R & Cit.\ F1 \\
\midrule
\textbf{Audio-direct} (MLLM ensemble)        & 593 & \textbf{0.853} & \textbf{0.679} & \textbf{0.947} & 0.672 & \textbf{0.786} \\
Text: Whisper $\rightarrow$ Gemini           & 593 & 0.818 & 0.620 & 0.775 & \textbf{0.769} & 0.772 \\
Text: Whisper $\rightarrow$ TF-IDF (sup.)     & 593 & 0.786 & 0.546 & 0.761 & 0.681 & 0.718 \\
\bottomrule
\end{tabular}
\end{table*}

\begin{figure}[t]
\centering
\includegraphics[width=\columnwidth]{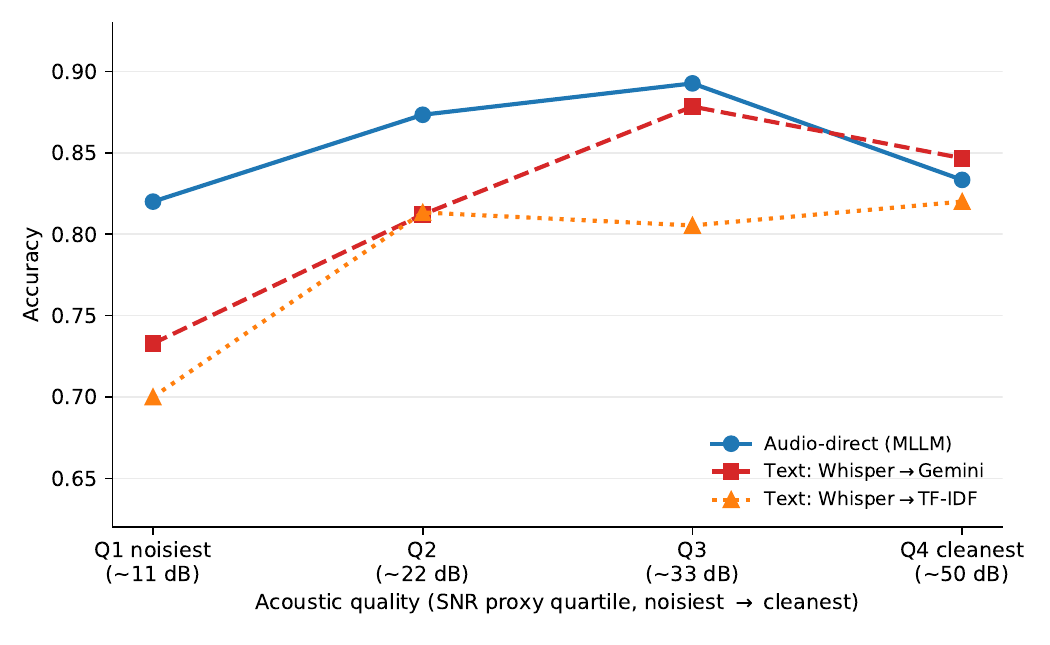}
\caption{Classification accuracy by acoustic quality (SNR-proxy quartile, noisiest to cleanest), for the audio-direct pipeline and the two transcription-based pipelines ($N = 593$, ${\sim}150$ speakers per quartile). The text pipelines match audio on clean speech (Q4) but degrade sharply on noisy speech (Q1), where ASR fails; the audio pipeline is comparatively flat.}
\label{fig:asr_robustness}
\end{figure}

\paragraph{Where audio earns its advantage.} The overall gap masks a sharp interaction with audio quality. Figure~\ref{fig:asr_robustness} and Table~\ref{tab:asr_robustness} split the $593$ speakers into quartiles of acoustic quality. On the cleanest audio (Q4), the text pipelines \emph{match or slightly exceed} the audio pipeline---unsurprising, since speaker role is heavily lexically marked (``my name is, I live at\ldots'' vs.\ procedural language), and on clean speech the transcript captures it. But as audio quality degrades, the text pipelines fall off sharply while the audio pipeline holds: on the noisiest quartile (Q1, ${\sim}11$\,dB), accuracy drops to $0.733$ (Gemini) and $0.700$ (TF-IDF) against $0.820$ for audio---a gap of $9$--$12$ points. This is the expected signature of ASR error, which is concentrated on exactly the noisy, overlapping, accented, and far-microphone speech that characterizes public comment \citep{koenecke2020racial}. Because real meeting corpora are full of such clips, the audio-direct approach is more robust where it matters and avoids inheriting transcription's failure modes.

\begin{table*}[t]
\centering
\caption{Accuracy by acoustic-quality quartile ($N \approx 150$ speakers each)}
\label{tab:asr_robustness}
\begin{tabular}{lcccc}
\toprule
SNR quartile (mean dB) & Audio-direct & Whisper$\to$Gemini & Whisper$\to$TF-IDF \\
\midrule
Q1 noisiest ($\sim$11\,dB)  & \textbf{0.820} & 0.733 & 0.700 \\
Q2 ($\sim$23\,dB)           & \textbf{0.873} & 0.812 & 0.813 \\
Q3 ($\sim$34\,dB)           & \textbf{0.893} & 0.878 & 0.805 \\
Q4 cleanest ($\sim$51\,dB)  & 0.833 & \textbf{0.847} & 0.820 \\
\bottomrule
\end{tabular}
\end{table*}

\noindent In sum, for classifying speaker roles, transcription is not a lossless intermediate step. Audio-direct classification matches the text pipeline on clean speech and is substantially more robust on the degraded audio that dominates real public-meeting recordings---while also avoiding the cost, latency, and documented racial and accent bias of ASR. A definitive version of this comparison, on all $1{,}761$ speakers and with a larger ASR model (which would give the text arm its strongest form), is left for the final draft.

\section{Theoretical Foundations}
\label{sec:appendix_theory}

This appendix provides formal grounding for three design choices in the pipeline: iterative bootstrapping as an optimization procedure, the sliding-window architecture as a 1D convolution analogy, and the ensemble rules as applications of the Condorcet jury theorem.

\subsection{Bootstrapping as Iterative Refinement}

The bootstrapping procedure (Algorithm~\ref{alg:bootstrap}) can be interpreted as a fixed-point iteration over the label space.  Let $\mathcal{L} = \{0,1\}^K$ denote the space of binary label assignments for all $K$ speakers in a meeting, and let $\Phi\colon \mathcal{L} \to \mathcal{L}$ denote the update operator that, given a current label assignment $\mathbf{L}_{t-1}$, produces a new assignment $\mathbf{L}_t$ by classifying each speaker with temporal exemplars drawn from $\mathbf{L}_{t-1}$.  Bootstrapping iterates $\mathbf{L}_t = \Phi(\mathbf{L}_{t-1})$ until convergence.

\textbf{Convergence guarantee.}  Because the label space $\mathcal{L}$ is finite ($|\mathcal{L}| = 2^K$) and the classifier $\Phi$ is deterministic (temperature $= 0$), the sequence $\{\mathbf{L}_t\}$ must either converge to a fixed point $\mathbf{L}^* = \Phi(\mathbf{L}^*)$ or enter a cycle.  In practice, we observe rapid convergence: accuracy rises over the first passes and then plateaus (84.8\% $\rightarrow$ 86.4\% $\rightarrow$ 86.9\% $\rightarrow$ 86.7\%) and the number of changed labels decreases, consistent with contraction toward a fixed point.

This structure is analogous to the expectation--maximization (EM) algorithm: Pass~0 provides an initial estimate (analogous to the E-step initialization), and subsequent passes refine labels using pseudo-labels as soft constraints (analogous to alternating E- and M-steps).  Unlike EM, however, we lack a formal likelihood objective, so convergence to a global optimum is not guaranteed.  The empirical monotonicity and rapid stabilization (within $T = 3$ passes) suggest that the landscape is well-behaved for this task.

\subsection{Sliding-Window Classification as 1D Convolution}

The sliding-window classifier (Section~\ref{sec:classification}) draws a formal analogy to one-dimensional convolutional neural networks.  Let $\mathbf{s} = [s_1, s_2, \ldots, s_N]$ denote the temporal sequence of speaker segments.  A 1D convolutional filter of width $w$ with stride~1 computes a local output at each position $i$:
\[
    h_i = \sigma\!\left(\sum_{j=0}^{w-1} W_j \cdot s_{i+j} + b\right), \quad i = 1, \ldots, N - w + 1,
\]
where $W_j$ are learned filter weights, $b$ is a bias term, and $\sigma$ is a nonlinear activation.  In our architecture, each LLM ``agent'' operates on a window $[s_{i-1}, s_i, s_{i+1}]$ (width $w = 3$, stride~1) and produces a classification for the center position $s_i$.  The agent's ``filter'' is the LLM itself: a fixed, pretrained function that maps three audio segments to a binary label.

The analogy is structural rather than mechanistic.  Unlike a learned CNN filter, the LLM applies in-context reasoning over the audio window rather than optimizing filter weights via backpropagation.  However, the architectural parallels are informative: (i)~overlapping receptive fields ensure that every segment's classification is informed by its immediate neighbors, enabling detection of local turn-taking patterns (official$\,\rightarrow\,$public$\,\rightarrow\,$official); (ii)~the stride-1 design provides translation equivariance over the temporal sequence, so the classification rule is position-invariant; and (iii)~the fixed window size constrains the model to exploit local structure, complementing the global temporal context provided by bootstrapping.

\subsection{Ensemble Theory: The Condorcet Jury Theorem}

The majority-vote ensemble is motivated by the Condorcet jury theorem \citep{condorcet1785}.  In its classical form, the theorem states: if each of $M$ independent voters has probability $p > 0.5$ of making the correct binary decision, then the probability that a majority votes correctly is
\[
    P_{\text{maj}} = \sum_{k=\lceil M/2 \rceil}^{M} \binom{M}{k} p^k (1-p)^{M-k},
\]
and $P_{\text{maj}} \to 1$ as $M \to \infty$.

In our setting, $M = 3$ classifiers with individual accuracies $p_1 = 0.847$ (5-shot), $p_2 = 0.867$ (Pass~3), and $p_3 = 0.852$ (window~2b).  The independence assumption is violated---all three classifiers use the same underlying MLLM---but the classifiers make \emph{complementary} errors due to different input representations (fixed exemplars vs.\ temporal exemplars vs.\ sliding window).  Empirically, pairwise inter-method $\kappa$ ranges from 0.725 to 0.751, and 24--30\% of each method's errors are unique (not shared with either other method), confirming substantial but imperfect correlation.  The empirical majority-vote accuracy (87.7\%) exceeds all individual classifiers, consistent with the Condorcet prediction under weak positive correlation.

The AND and OR ensembles represent principled tradeoffs along the precision--recall frontier:
\begin{itemize}[nosep]
    \item \textbf{AND rule} (predict public participant only if \emph{both} agree): increases the effective decision threshold, yielding high precision at the cost of recall.  For two classifiers with independent false-positive rates $\alpha_a$ and $\alpha_b$, the AND rule achieves false-positive rate $\alpha_a \cdot \alpha_b < \min(\alpha_a, \alpha_b)$.
    \item \textbf{OR rule} (predict public participant if \emph{either} agrees): decreases the threshold, yielding high recall at the cost of precision.  The OR rule achieves false-negative rate $\beta_a \cdot \beta_b < \min(\beta_a, \beta_b)$ under independence.
\end{itemize}
\noindent These theoretical properties are borne out empirically: the AND ensemble of 5-shot $\cap$ window~2b achieves the highest public precision (0.930), while the OR ensemble of Pass~3 $\cup$ window~2b achieves the highest public recall (0.854), with the majority vote providing the best F1 (0.832).

\end{document}